\documentclass[preprint,nofootinbib,noshowkeys,a4paper]{revtex4}
\usepackage{graphicx} 
\usepackage{bm}

\def\slashchar#1{\setbox0=\hbox{$#1$}           % set a box for #1 
   \dimen0=\wd0                                 % and get its size
   \setbox1=\hbox{/} \dimen1=\wd1               % get size of /
   \ifdim\dimen0>\dimen1                        % #1 is bigger
      \rlap{\hbox to \dimen0{\hfil/\hfil}}      % so center / in box
      #1                                        % and print #1
   \else                                        % / is bigger
      \rlap{\hbox to \dimen1{\hfil$#1$\hfil}}   % so center #1
      /                                         % and print /
   \fi}                                         %

\begin{document}

\title{Transport Coefficients and Ladder Summation \\ 
in Hot Gauge Theories}
\author{Manuel A. \surname{Valle Basagoiti}}
\email{wtpvabam@lg.ehu.es}
\affiliation{Departamento de F\'\i sica Te\'orica,  
Universidad del Pa\'\i s Vasco, Apartado 644, 
E-48080 Bilbao, Spain}
\date{April 2002}

\begin{abstract}
     
    We show how to compute transport coefficients in gauge theories 
    by considering the expansion of the Kubo formulas in terms of ladder
    diagrams in the imaginary time formalism.  All summations over
    Matsubara frequencies are performed and the analytical continuation to
    get the retarded correlators is done.  As an illustration of the
    procedure, we present a derivation of the transport equation for the
    shear viscosity in the scalar theory.  Assuming the Hard Thermal Loop
    approximation for the screening of distant collisions of the hard
    particles in the plasma, we derive a couple of integral equations for
    the effective vertices which, to logarithmic accuracy, are shown to be
    identical to the linearized Boltzmann equations previously found by
    Arnold, Moore and Yaffe.
   
\end{abstract}

\maketitle

%%%%%%%%%%%%%%%%%%%%%%%%%%%%%%%%%%%%%%%%%%%%%%%%%%%%%%%%%%%%%%%%%%%%%%%%%%%%%%%%%%

\section{\label{intro}Introduction}

The development of a transport theory  for QCD in the regime of 
high temperature
%of transport properties in gauge theories at high temperature 
has turned out a valuable pursuit with the advent of heavy ion  colliders 
which provide novel tools for the study of the properties of highly excited 
matter.
From a purely theoretical point of view, the computation of transport 
coefficients amounts a challenge even in weakly coupled theories,  because these 
quantities usually depend non-analytically on the coupling constant.
%of the non-analytical dependence on the coupling constant. 
In most 
previous computations~\cite{Baym,Heiselberg,Baym2}, 
a kinetic approach based in the Boltzmann equation 
has been used. 
It is only recently within this framework that a reliable computation to 
logarithmic accuracy in gauge theories has been reported~\cite{Arnold1}. 
The complete leading order is still unavailable except for the case 
of a gauge theory with a large number of fermionic species~\cite{Moore}.  

However, there exists an alternative approach based on Kubo formulas for  
appropriate correlation functions, which has 
been largely used in the context of low energy many-body physics~\cite{Mahan}. 
For the electrical and thermal conductivities of ordinary metals and superconductors, 
the computation of the current correlators 
requires the resummation of an infinite class of ladder diagrams, a task that 
in the complete relativistic setting of gauge theories  
usually appears as a very hard  
issue, partly motivating the scant of this approach.
In relativistic transport theory, the resummation of ladder diagrams has %only 
been performed for the scalar theory  
by Jeon~\cite{Jeon},
who proved the equivalence with the Boltzmann equation, 
and since repeated a few times~\cite{Carrington,Wang}.  
Also, a simplified ladder summation has been performed  
for the computation of the leading-log order of the color 
conductivity~\cite{Martinez}. Recently, Arnold, Moore and 
Yaffe~\cite{Arnold2,Arnold3} 
have performed a ladder summation in order to account for 
the effect of multiple scattering in the process of photon production 
from a QCD plasma. 

The purpose of this paper is twofold. First, we 
wish to explicitly show how to perform the summation of 
a restricted set of ladder 
diagrams in gauge theories within the imaginary time formalism 
of thermal field theory.
In contrast to the work of Jeon~\cite{Jeon}, 
we will not use the series of cut 
ladder diagrams.
Rather, we closely follow a treatment due to Holstein~\cite{Holstein}
who, a long time ago, performed a ladder summation in order to compute 
the transport properties of the low-energy electron-phonon gas. 
In this approach, 
one first identifies the required analytic continuation of the 
effective vertex function 
entering in the current correlator, and then  writes the integral equation 
for this vertex, summing all ladders. As shown below, this approach exactly 
reproduces the correct transport equation for the shear viscosity in the 
scalar theory.  

On the other hand, we will try to derive the logarithmic 
accuracy of the transport coefficients by  only considering 
the role played by the soft degrees of freedom which are 
exchanged in the collisions between the plasma constituents. 
This requires the use of the Hard Thermal Loop (HTL) 
approximation~\cite{HTL} for the internal lines associated 
with the rungs  of the diagrams, 
and the introduction of an arbitrary momentum scale $q_c$ 
separating the hard and soft 
ranges of the momentum transfer~\cite{Yuan}. 
An important step towards the complete computation of the 
hard contribution  was already recently made by the authors 
of  Ref.~\cite{Arnold1}, who calculated the infrared 
logarithmic divergences of the collision terms of linearized 
Boltzmann equation written in terms of unscreened interactions.

Our main results are a couple of integral equations 
for the effective vertices, encoding the effects of distant collisions 
in the plasma coming from soft momentum transfer $q < q_c$. 
Although  these equations are necessarily incomplete, 
they reproduce the required logarithmic 
dependence on $q_c$ which makes possible the 
eventual cancellation of the arbitrary scale $q_c$ in the final 
result. 
Hence, they reproduce the known results~\cite{Arnold1} for the transport 
coefficients to logarithmic accuracy. 

The plan of this paper is as follows. In Sec.~\ref{sec:basic}, we review  
some standard material on the imaginary time formalism of thermal field 
theory and Kubo formulas. Here, we include a useful 
summation formula over Matsubara frequencies and the procedure of analytic
continuation.  In Sec.~\ref{sec:phi4} we show how to derive the shear
viscosity of $\lambda \phi^4$ theory.  Section~\ref{sec:soft} deals
with the simplifications which appear by summing the ladders when only
the effect of soft momentum transfer is considered.  
Then, we derive the
corresponding contribution to the transport equations for the electrical
conductivity and the shear viscosity.  Section~\ref{sec:log} presents  a
brief derivation based on sum rules of the logarithmic terms in the
transport coefficients, and Sec.~\ref{sec:end} closing the paper
contains a summary and prospects.  There are short appendixes
with some details about spectral densities, sum rules and the relevant thermal
widths to be included in the propagators of hard particles.
 
%%%%%%%%%%%%%%%%%%%%%%%%%%%%%%%%%%%%%%%%%%%%%%%%%%%%%%%%%%%%%%%%%%%%%%%%%%%%%%%%%%%%%%%%%

\section{\label{sec:basic}Basic formalism}
\subsection{Single particle spectral densities}

The basic element of a diagram in the Imaginary-Time Formalism is the
Matsubara propagator depending on the purely imaginary frequencies $i
\omega_{n}= i \pi n/\beta$ (with $n$ even for bosons and odd for
fermions),
\begin{equation}\label{eq:lehmann}
\Delta(i\omega_{n},{\mathbf q}) =
\int_{-\infty}^{\infty}\frac{d q^0}{2\pi} \frac{\rho(Q)}{q^0-i\omega_{n}},
\end{equation} 
where the real quantity $\rho(Q)$ (with $Q^\mu=(q^0,{\mathbf q})$,
$q^0$ real) is the single particle spectral density. 
The analytical continuation $i\omega_{n} \rightarrow z$ defines a 
function $\Delta(z,{\mathbf q})$ of a complex variable $z$ 
which is analytical off the real axis and the discontinuity 
through the branch cut along $\rm{Im}\,z=0$ is 
proportional to $\rho(Q)$. 
The different Green's
functions for real frequency can be constructed from the spectral
density.  For instance, the bosonic Wightman functions
$\Delta^{>,\,<}(Q)$ are given by 
\begin{equation}
\Delta^{>}(Q)=\left(1+n_{b}(q^0) \right) \rho(Q), \qquad
\Delta^{<}(Q)=n_{b}(q^0) \rho(Q),
\end{equation}
where $n_{b}(q^0)=1/(e^{\beta q^0}-1)$ is the bosonic occupation
number.  The retarded and advanced Green's functions, which will play
an important role in our discussion, are
\begin{equation}
\Delta^{\rm{ret}}(Q)=\Delta(q^0+ i 0^+,{\mathbf q}), \qquad 
\Delta^{\rm{adv}}(Q)=\Delta(q^0- i 0^+,{\mathbf q}), 
\end{equation}
and $\rho(Q)=2\,\rm{Im}\,\Delta^{\rm{ret}}(Q)$. 

For a free particle, the spectral density is given by a superposition
of delta functions, with support on the mass shell,
$p_{0}^2=\varepsilon_{p}^2$.  If the interactions are weak, a delta function
can be replaced by a Lorentzian with a small width $\Gamma_{p}$, related
to the imaginary part of the energy at the poles of the retarded propagator, 
$p^0 = \pm \varepsilon_{p}- i \Gamma_{p}$, 
\begin{equation}
       2 \pi \delta(p^0 \mp \varepsilon_{p}) \longmapsto 
       \frac{2 \Gamma_{p}}{(p^0\mp\varepsilon_{p})^2+\Gamma_{p}^2} ,
\end{equation}
which gives rise to an analytical propagator
\begin{equation}
\Delta(z,{\mathbf p}) = \int_{-\infty}^{\infty}\frac{d p^0}{2\pi} 
\frac{2 \Gamma_{p}}{(p^0 \mp \varepsilon_{p})^2+\Gamma_{p}^2}\,
\frac{1}{p^0-z} = \frac{-1}{z \mp \varepsilon_{p} +
i \Gamma_{p}\,\rm{sign}(\rm{Im}\,z)} .
\end{equation}
Thus, for this case, the function $\Delta(z,{\mathbf p})$ has a branch cut at real
axis in the complex $z$-plane and it has no poles.  
 
At high temperature, the particles entering in the collision processes
occurring in the plasma are mostly  particles propagating nearly on-shell 
with hard momentum, $P \sim T$.  Their spectral densities may be
approximated by a combination of two Lorentzians
\begin{eqnarray}\label{eq:bose}
    \rho_{b}(P) &=&
\frac{1}{p}\left(\frac{\Gamma_{p}}{(p^0-p)^2+\Gamma_{p}^2} -
\frac{\Gamma_{p}}{(p^0+p)^2+\Gamma_{p}^2}\right), \\
\label{eq:fermi}
\rho_{f}(P) &=& \left(
\frac{2\gamma_{p}}{(p^0-p)^2+\gamma_{p}^2}h_{+}(\hat{\mathbf{p}})+
\frac{2\gamma_{p}}{(p^0+p)^2+\gamma_{p}^2}h_{-}(\hat{\mathbf{p}})\right),
\end{eqnarray}
where $ h_{\pm}(\hat{\mathbf{p}}) = 
(\gamma^0 \mp {\bm{\gamma}} \cdot \hat{\mathbf{p}} )/2$, and the thermal 
widths $\Gamma_{p}$ and $\gamma_{p}$ are the imaginary parts of 
the transverse piece of the on-shell gluon self-energy and the quark,
respectively
\begin{eqnarray}
    \Gamma_{p} &=& -\frac{1}{2 p}{\rm Im}\,
    \Pi_{\rm{T}}^{\rm{ret}}(p^0=p,\mathbf{p}), \\
    \gamma_{p} &=& -\frac{1}{4 p}{\rm{tr}} \left({\slashchar p}\,
    {\rm Im}\,\Sigma^{\rm{ret}}(p^0=p,\mathbf{p}) \right). 
\end{eqnarray}
The shift in the real part of the energy can be ignored since it is 
perturbatively small when the energy is $O(T)$. 

In gauge theories, the imaginary part of the thermal self-energies
receives contributions from various scattering processes which give a 
different dependence on the coupling constant.  
Generically, two body scattering
processes in which a soft bosonic excitation is exchanged yield a
parametric dependence at leading order as $\Gamma_{p}, \gamma_{p}
\propto g^2 T\log(\Lambda_{\rm{max}}/\Lambda_{\rm{min}})$,
whereas processes in which a soft fermionic excitation is exchanged
yield a parametric dependence $\Gamma_{p}, \gamma_{p} \propto g^4
T^2\log(q_{c}/g T)/p$.  The cutoff $q_{c}$ is a scale separating
semihard and hard momentum transfers, restricted by $g T \ll q_{c} \ll
T$ but otherwise arbitrary, and $\Lambda_{\rm{max}}$ can be chosen
of order $g T$.  As will be explicitly showed below, the infrared
sensitivity to the lower cutoff $\Lambda_{\rm{min}} \sim g^2 T$
entirely disappears from the transport coefficients to be computed.

On the other hand, the temperature Green's functions for the soft bosonic and
fermionic excitations are determined by the single particle spectral
densities in the hard thermal loop approximation, ${}^\ast
\rho_{\rm{L,T}}(\omega,{\mathbf q})$ and ${}^\ast
\Delta_{\pm}(\omega,{\mathbf q})$, 
respectively~\cite{HTL,Blaizot,Lebellac}.   These are presented
in the appendices~\ref{app:srule} and~\ref{app:dampings}.  
Already, let us note here that only the Landau
damping piece of these will contribute to the screening of distant
collisions.

\subsection{\label{kubo}Kubo formulas and the ladder approximation}

Our starting point is the Kubo formula expressing a given transport
coefficient in terms of the low frequency, zero momentum
limit of the spectral density for the corresponding correlation
function.  For the electrical conductivity and the shear viscosity the
formulas are~\cite{Mahan,Jeon}
\begin{eqnarray}\label{eq:cond}
	\sigma &=& \frac{1}{6}\lim_{\omega \rightarrow 0}\,
	           \lim_{{\mathbf q}\rightarrow 0}
	           \frac{\partial}{\partial \omega}
	           \rho_{J J}(\omega,{\mathbf q}) ,  \\ 
\label{eq:sv}
	\eta &=& \frac{1}{20}\lim_{\omega \rightarrow 0}\,
	           \lim_{{\mathbf q}\rightarrow 0}
	           \frac{\partial}{\partial \omega}
	           \rho_{\pi \pi}(\omega,{\mathbf q})  ,           
\end{eqnarray}
where, as usual,  the spectral densities are related to the Fourier transform of
the retarded correlators by
\begin{eqnarray}
	\rho_{J J}(\omega,{\mathbf q})&=& 2\,{\rm Im} 
	              \int_{-\infty}^{\infty}dt \int d^3x\, 
	               e^{i\omega t-i {\mathbf q}\cdot {\mathbf x}}
	               \langle\left[J_i(t,{\mathbf x}), J_k(0)\right]\rangle 
\theta(t) \delta_{i k}, \\
\rho_{\pi \pi}(\omega,{\mathbf q})&=& 2\,{\rm Im} 
	               \int_{-\infty}^{\infty}dt \int d^3x\, 
	               	e^{i\omega t-i {\mathbf q}\cdot {\mathbf x}}
	                \langle\left[\pi_{ij}(t,{\mathbf x}), 
	                \pi_{ij}(0)\right]\rangle \theta(t),
\end{eqnarray} 
and the averages are evaluated in the equilibrium grand canonical 
ensemble. An  efficient way to compute a retarded correlator 
$\Pi_{AA}^{\rm{ret}}(\omega,{\mathbf q})$ (with $A$ denoting
collectively the indices of the appropriate current) is to exploit the 
spectral representation for complex frequency $z$.  This  provides a
direct connection with the temperature Green's function $\Pi_{AA}(i
\nu_n,{\mathbf q})$, via analytic continuation $i \nu_n \rightarrow
\omega+ i 0^+$.  Thus, a first step in our basic task is to evaluate
$\Pi_{AA}(i \nu_n,{\mathbf 0})$ within the imaginary time formalism.

After the laborious diagrammatic analysis explicitly performed
for the scalar theory~\cite{Jeon}, 
the conclusion is that, in order to account for all 
leading-order contributions to the shear viscosity, 
a set of uncrossed ladder diagrams must be summed.
On the other hand, for the processes of photon production 
from a QCD plasma, the authors of Refs.~\cite{Arnold2,Arnold3}
have developed detailed power counting arguments which enforce 
the resummation of the
uncrossed ladder graphs made of gauge boson rungs. 

A key point for understand
the equal footing of this class of diagrams is the presence of pairs
of propagators carrying nearly the same momenta, which leads to a
dependence $1/\Gamma_p$ (or $1/\gamma_p$) by each such a propagator
pair.  Hence, it is clear that a $(n+1)$-loop diagram with $n$
uncrossed rungs leads to a dependence proportional to $\alpha^2
(1/\Gamma_p)^{n+1} (g^2)^n$, where the first $\alpha^2$ comes from the
two external insertions, and each rung introduces a factor $g^2$.  The
derived result in both cases~\cite{Jeon} and~\cite{Arnold2} is a linear
integral equation for an effective vertex function, which is
completely equivalent to the linearized transport equation for the
problem.  Here, we will proceed by assuming the dominance of the same
set of ladder diagrams, relying on a posteriori check of its
consistency.

\begin{figure}
    \includegraphics{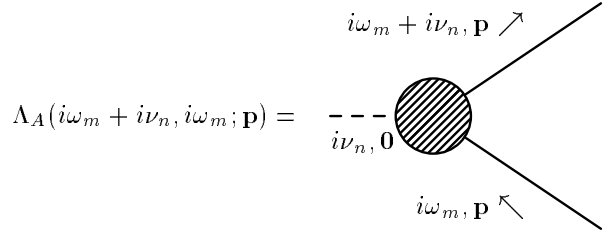}
    \caption{\label{fig:blob} Labels of the effective vertex.}
\end{figure}

At this point, let us then to introduce an amputated effective vertex 
$\Lambda_{A}(i\omega_m + i \nu_n,i\omega_m ; {\mathbf p})$, associated with 
two hard external bosonic or fermionic lines $(i\omega_m + i \nu_n,{\mathbf p})$ ,  
$(i\omega_m,{\mathbf p})$ and with an insertion of zero external momentum 
but non-zero frequency   
$(i\nu_n, {\mathbf 0})$, corresponding to the appropriate current ($J_i$
or $\pi_{ij}$).  This effective vertex (Fig.~\ref{fig:blob}) is the sum of all
vertices, each one of them with a number $n$ of rungs associated with
the exchanged excitations, and presumably will encode all collision
effects at leading order.  The vertex having zero rungs,
$\Lambda_{A}^{(0)}({\mathbf p})$, does not depend on the frequencies
but can depend on the momentum ${\mathbf p}$.
%%%%%%%%%%%%%%%%%%%%%%%%%%%%%%%%%%%%%%%%%%%%%%%%%%%%%%%%%%%%%%
%%%%%%%%%%%%%%%%%%%%%%%%%%%%%%%%%%%%%%%%%%%%%%%%%%%%%%%%%%%%%%%

\subsection{Summation over Matsubara frequencies and analytic continuation}

Let us now to examine the summation over Matsubara frequencies, which
generically is  involved in the evaluation of a temperature correlator
$\Pi_{AA} (i\nu_n,{\mathbf 0})$,
\begin{equation}\label{eq:all}
  T \sum_{\omega_m} G(i \omega_m +i \nu_n, {\mathbf p})
  \Lambda_A(i\omega_m + i \nu_n,i\omega_m ; {\mathbf p}) 
                  G(i \omega_m, {\mathbf p}) \Lambda_A^{(0)}({\mathbf p}) ,  
\end{equation}
where the Matsubara propagators have spectral densities of the
form~(\ref{eq:bose}) or (\ref{eq:fermi}).

To one-loop order, the effective
vertex reduces to $\Lambda_A^{(0)}$ and the above sum over frequencies
becomes
\begin{equation}\label{eq:oneloop}
H(i \nu_n,{\mathbf p})=
T \sum_{\omega_m} G(i \omega_m +i \nu_n, {\mathbf p}) 
                  G(i \omega_m, {\mathbf p}),
\end{equation} 
with $\nu_n$ even in any case.  Now, the function to be summed is a
product of Green's functions and one may proceed by expressing the
Green's functions in terms of their spectral representations given by
Eq.~(\ref{eq:lehmann}).  With this replacement, the resulting expression
involves a double frequency integral of the product resulting of the
spectral densities and the elementary sum
\begin{equation}
      T \sum_{\rm{even,\,odd}\, m} \frac{1}{i \omega_m - \omega_1}
       \frac{1}{i \omega_m+i\nu_n - \omega_2} = 
      \mp\, \frac{n_{b,f}(\omega_1)-n_{b,f}(\omega_2)}
       {i \nu_n+\omega_1-\omega_2},
\end{equation}
where $n_{b,f}$ denote boson or fermion occupation factors. 
However, it is more convenient to use an alternative procedure based on 
contour integration in the $z$-plane of an appropriate function. 

\begin{figure}
    \includegraphics{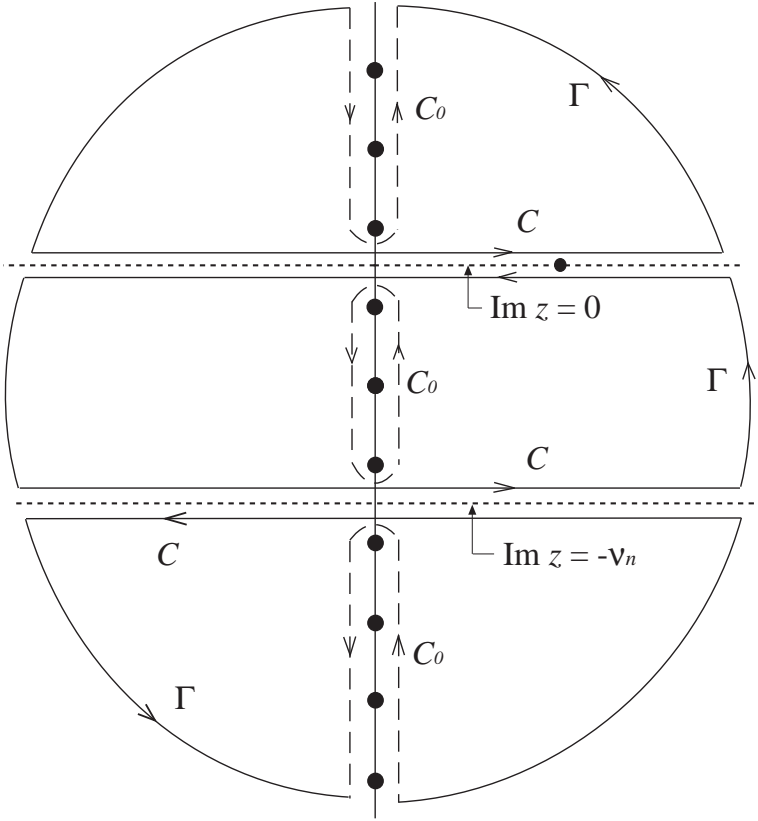}
    \caption{\label{fig:contour}Integration contour for the sum 
    $T \sum_m F(i\omega_m)$.  The original contour $C_0$ consists of three
    dashed circuits encircling the Matsubara frequencies.  The deformed
    contour $C$ goes along both sides of the branch cuts.  A possible pole
    of $F(z)$ at the real axis is depicted for its relevance in the
    summation of the vertex equation.}
\end{figure}

This procedure was used, a long time ago, by
Holstein~\cite{Holstein} for to derive diagrammatically the transport
properties of an electron-phonon gas.  Here, we closely follow the
treatment of Holstein.
To perform the summation~(\ref{eq:oneloop}), we consider the function
$G(z+i \nu_n, {\mathbf p})G(z, {\mathbf p})n_{b,f}(z)$, and a contour
integration $C_0$ made of three circuits enclosing the poles of $n_{b,f}(z)$
at the imaginary axis but avoiding the other possible poles of $G G$
(in this case there are absent), and the two branch cuts at 
${\rm Im}\,z=0$ and ${\rm Im}(z+i \nu_n)=0$. This contour may be deformed to 
$C$ and $\Gamma$ as shown in Fig.~\ref{fig:contour}. 
The contributions from the
large arcs $\Gamma$ vanish and we are left with the integrals along $C$.  
Then, Eq.~(\ref{eq:oneloop}) becomes after summation
\begin{eqnarray}
H(i \nu_n,{\mathbf p})&=&\pm \frac{1}{2\pi i} 
\int_C dz\, G(z+i \nu_n, {\mathbf p})G(z, {\mathbf p})n_{b,f}(z)\nonumber \\
&=& \mp\int_{-\infty}^{\infty}\frac{d\xi}{2\pi i} n_{b,f}(\xi)\, 
  \left\{ \left[G(i\nu_n+\xi,{\mathbf p})+
                G(-i\nu_n+\xi,{\mathbf p})\right] 
                G^{\rm adv}(\xi,{\mathbf p})\right. \nonumber \\
&&-\left. G^{\rm ret}(\xi,{\mathbf p})
          \left[G(i\nu_n+\xi,{\mathbf p})+
                G(-i\nu_n+\xi,{\mathbf p})\right]\right\}
\end{eqnarray}
where $\xi$ is a variable specifying the position at the branch cuts.
More generally, if the function $F(z)$ to be summed has poles in the complex
$z$-plane, the summation formula which will be extensively used in
what follows  is
\begin{equation}\label{eq:suma}
T \sum_{{\rm even,\, odd}\, m} F(i \omega_m)= 
\mp \sum_{\rm poles}n_{b,f}(z_i)
 {\rm Res}(F,z=z_i) \pm \sum_{\rm cuts}
 \int_{-\infty}^{\infty}\frac{d\xi}{2\pi i}n_{b,f}(\xi)
  {\rm Disc}F. 
\end{equation}
%%%%%%%%%%%%%%%%%%%%%%%%%%%%%%%%%%%%%%%%%%%%%%%%%%%%%%%%%%%%%%
%%%%%%%%%%%%%%%%%%%%%%%%%%%%%%%%%%%%%%%%%%%%%%%%%%%%%%%%%%%%%%%

Application of this formula to the summation in Eq.~(\ref{eq:all}) requires 
the determination of the singularities of $\Lambda_A(z + i \nu_n, z ; {\mathbf p})$.  
We argue that the only singularities are two branch cuts at the lines 
${\rm Im} ( z + i \nu_n) = 0$ and ${\rm Im} (z) = 0$. This is a consequence of 
the recurrence relation for the vertex with $n$ rungs,
\begin{eqnarray}
   &&\Lambda_A^{(n)}(i\omega_m + i \nu_n , i\omega_m; {\mathbf p}) 
   =
   %\Lambda_A^{(n-1)}(i\omega_m + i \nu_n , i\omega_m; {\mathbf p})
   %+
   T \sum_{\nu_q} \int \frac{d^4Q}{(2 \pi)^4}\,G(i \omega_m +i \nu_n +
   i\nu_q, {\mathbf p} + {\mathbf q}) \nonumber \\
   &&\qquad \times 
   \Lambda_A^{(n-1)}(i\omega_m + i \nu_n + i\nu_q, i\omega_m + i\nu_q; 
   {\mathbf p} + {\mathbf q}) 
   G(i \omega_m + i \nu_q, {\mathbf p} + {\mathbf q}) 
   \frac{\rho(Q)}{q^0 - i\nu_q}\,,
\end{eqnarray}
where $\rho(Q)$ is the spectral density corresponding to the added
rung and $(i\nu_q,{\mathbf q})$ label the momenta running through the
loop.  For $n=1$, it is clear that 
Eq.~(\ref{eq:suma}) implies that $\Lambda_A^{(1)}$ has only the
singularities of the product of the Green's functions.  Then, it
follows from mathematical induction that $\Lambda_A^{(n)}$, and hence
$\Lambda_A$, inherits the same property.

Now, we are ready to perform the summation in Eq.~(\ref{eq:all}) and the
subsequent analytic continuation.  Making use of Eq.~(\ref{eq:suma}) we
may write
\begin{eqnarray}
&&T \sum_{\omega_m} G(i \omega_m +i \nu_n)
  \Lambda_A(i\omega_m + i \nu_n,i\omega_m) 
                  G(i \omega_m)= 
                  \mp\int_{-\infty}^{\infty}\frac{d\xi}{2 \pi i} n_{b,f}(\xi)	
                  \nonumber \\ 
        && \quad \times \left\{ \left[ G(\xi+i \nu_n)
       \Lambda_A(\xi+i \nu_n, \xi-i 0^+)+ G(\xi-i \nu_n) 
       \Lambda_A(\xi-i 0^+,\xi-i \nu_n) \right]
        G^{\rm adv}(\xi) \right.  \nonumber \\
        && \quad -\left.  G^{\rm ret}(\xi)\left[ 
        \Lambda_A(\xi+i \nu_n, \xi+i 0^+)G(\xi+i \nu_n)+ 
        \Lambda_A(\xi+i 0^+, \xi-i \nu_n)G(\xi-i \nu_n) \right] \right\},
\end{eqnarray}
where the dependence on $\mathbf p$ is not explicitly exhibited.  Next, the
analytical continuation $i \nu_n \rightarrow \omega + i 0^+$ of the
previous expression yields
\begin{eqnarray}
	&&\mp \int_{-\infty}^{\infty}\frac{d\xi}{2 \pi i}
	\left\{n_{b,f}(\xi+\omega) 
	     G^{\rm adv}(\xi+\omega,{\mathbf p}) 
	     \Lambda_A(\xi+\omega-i 0^+, \xi-i 0^+; {\mathbf p})
	     G^{\rm adv}(\xi,{\mathbf p}) \right.\nonumber \\*
	&&\qquad\qquad -n_{b,f}(\xi) 
	     G^{\rm ret}(\xi+\omega,{\mathbf p})  
	     \Lambda_A(\xi+\omega+i 0^+, \xi+i 0^+; {\mathbf p})
	     G^{\rm ret}(\xi,{\mathbf p})  \nonumber  \\* 
	&&-\left.\left[n_{b,f}(\xi+\omega)-n_{b,f}(\xi)\right]
	     G^{\rm ret}(\xi+\omega,{\mathbf p}) 
	     \Lambda_A(\xi+\omega+i 0^+, \xi-i 0^+; {\mathbf p})
	     G^{\rm adv}(\xi,{\mathbf p}) \right\},
\end{eqnarray}
where we have rearranged some terms  by a shift of the integration variable. 
At this point,  a simplification arises since the integrand of the above
expression contains a large term coming from the product $G^{\rm
ret} G^{\rm adv}$. This is due to the fact that the two pairs of poles of
this product are located at both sides on real axis in the
$\xi$-plane.  Thus, in the limit $\omega \rightarrow 0$, the
contribution of $G^{\rm ret} G^{\rm adv}$ to the integral is
inversely proportional to the distance between the poles given by the
thermal width.  The other products $G^{\rm ret} G^{\rm ret}$
and $G^{\rm adv} G^{\rm adv}$ make a much smaller
contribution, due to the cancellation between the residues at the
poles.

Therefore, noting that the effective
vertex $\Lambda_A(\xi+i 0^+, \xi-i 0^+; {\mathbf p})$ is a real 
quantity\footnote{This follows from mathematical induction, taking into
account that the zero-order vertex $\Lambda_A^{(0)}({\mathbf p})$ is
real.}, and $n_{b,f}'(\xi)=-\beta n_{b,f}(\xi) (1\pm n_{b,f}(\xi))$,
the required zero frequency slope of the retarded correlator may be
written as
\begin{eqnarray}\label{eq:slope}
	\left.\frac{\partial} {\partial\omega}
	{\rm Im}\, \Pi_{AA}^{\rm ret}(\omega,{\mathbf 0})\right|_{\omega=0}
	&= & \zeta\,  \beta \int \frac{d^3{\mathbf p}}{(2\pi)^3}
	\Lambda_A^{(0)}({\mathbf p}) \int_{-\infty}^{\infty}\frac{d\xi}{2 \pi}
	n_{b,f}(\xi) \left(1\pm n_{b,f}(\xi)\right) \nonumber \\
           &&\times G^{\rm ret}(\xi,{\mathbf p}) 
           \Lambda_A(\xi+i 0^+, \xi-i 0^+; {\mathbf p})
	   G^{\rm adv}(\xi,{\mathbf p}), 
\end{eqnarray}
where the prefactor $\zeta$ accounts for the symmetric factor
associated with the one-loop diagram corresponding to $\Pi_{AA}$. 
This factor is $1/2$ or $1$ depending on whether the $G$ functions correspond to a
self-conjugate field or not.  
It is important to notice the correspondence
between Eq.~(\ref{eq:slope}) and the expression for a transport
coefficient in terms of $\delta n_{A,\pm}({\mathbf p})$, denoting the
departure from equilibrium of the single particle (anti-particle)
density function.  Anticipating a contribution of $G^{\rm ret}
G^{\rm adv} \propto \sum_{\pm}\delta(\xi \mp \left|{\mathbf
p}\right|)/\Gamma_p$, such a correspondence is clear with the
appropriate identification
\begin{equation}
       \delta n_{A,\pm}({\mathbf p})\propto 
       \frac{1}{\Gamma_p} 
       \Lambda_A(\pm\left|{\mathbf p}\right|+i 0^+, 
                 \pm\left|{\mathbf p}\right|-i 0^+; {\mathbf p}) . 
\end{equation}

%%%%%%%%%%%%%%%%%%%%%%%%%%%%%%%%%%%%%%%%%%%%%%%%%%%%%%%%%%%%%%%%%%%%%%%%%%%%%

\section{\label{sec:phi4}The transport equation for shear 
viscosity in $\lambda \phi^4$ theory}

With the purpose of making clear the basic procedure to be used in gauge theories,
we present a simple alternative derivation of the transport equation for 
the scalar theory in the form previously obtained by Jeon~\cite{Jeon}. 
The starting point is the 
linear integral equation for the effective vertex before analytic continuation
\begin{eqnarray}\label{eq:svertex}
	\Lambda_{ij}(i\omega_m &+& i \nu_n , i\omega_m; {\mathbf p}) = 
	 \Lambda_{ij}^{(0)}({\mathbf p})+\frac{1}{2} 
   T \sum_{\nu_q} \int \frac{d^4Q}{(2 \pi)^4}\,
   G(i \omega_m +i \nu_n + i\nu_q, {\mathbf p} + {\mathbf q}) \nonumber \\  
   &&\times 
   \Lambda_{ij}(i\omega_m + i \nu_n + i\nu_q, i\omega_m + i\nu_q; 
   {\mathbf p} + {\mathbf q}) 
   G(i \omega_m + i \nu_q, {\mathbf p} + {\mathbf q}) 
   \frac{\rho(Q)}{q^0 - i\nu_q}\, ,
\end{eqnarray}
where the factor $1/2$ is due to combinatorics, and
$\Lambda_{ij}^{(0)}({\mathbf p}) = 2 p^2 ({\hat p}^i {\hat p}^j -
\delta^{i j}/3)$\footnote{
The factor $2$ is due to the sum over the two permutations of the momenta 
corresponding of the scalar field.}.  This equation is shown in
Fig.~\ref{fig:phi4}.  The spectral density $\rho(Q)$ associated with
the bubble in the rung is given by
\begin{equation}
\rho(q^0, {\mathbf q})= 2\,{\rm Im}\, \left.\left\{\lambda^2 T
\sum_{\nu_k} \int \frac{d^3{\mathbf k}}{(2 \pi)^3}
\frac{1}{(\nu_q+\nu_k)^2 + \left|{\mathbf k} + {\mathbf q}\right|^2}\,
\frac{1}{\nu_k^2 + {\mathbf k}^2}\right\} \right|_{i\nu_q=q^0+i 0^+}\,.
\end{equation}
By performing the previous sum (see appendix~{\ref{app:bubble}), and using 
the expression for the free Wightman function, 
\begin{equation}
G_0^>(P) = %\left(1 + n_b(p^0)\right) \rho_0(P) =  
\left(1 + n_b(p^0)\right) 2\pi\, {\rm sgn}(p^0)\, 
\delta(p_0^2-{\mathbf p}^2) ,
\end{equation}
one may write the spectral density for
the rung in a symmetric form 
\begin{equation}\label{eq:bubble}
\rho(Q)=\frac{\lambda^2}{n_b(q^0)}
   \int \frac{d^4 K}{(2 \pi)^4}\, G_0^>(-Q-K) G_0^>(K).  
\end{equation}
The relation of this spectral density with the kernel $L_{\rm Boltz}$ in 
the treatment of Jeon~\cite{Jeon} is therefore
\begin{equation}\label{ident}
\rho(Q)=\frac{2}{n_b(q^0)} L_{\rm Boltz}(-Q).
\end{equation}
Note that $\rho(Q)$ is odd in $q^0$. 

%%%%%%%%%%%%%%%%%%%%%%%%%%%%%%%%%%%Figure%%%%%%%%%%%%%%%%%%%%%%%%%%%%%%%%%%%
\begin{figure}
    \includegraphics{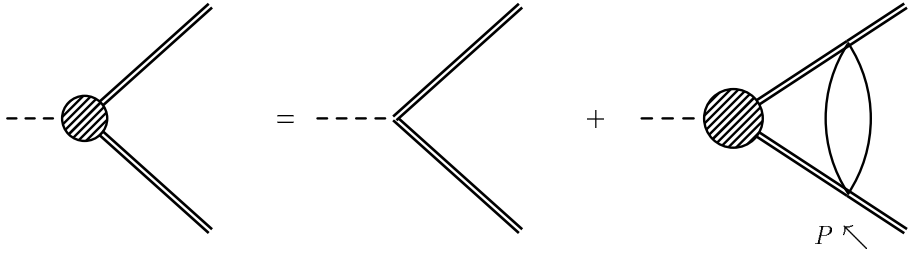} 
    \caption{\label{fig:phi4}Equation for the
    effective vertex in the $\lambda \phi^4$ theory.  The double lines
    represent the propagators of the hard particles including the
    $O(\lambda^2)$ thermal width.}
\end{figure}
%%%%%%%%%%%%%%%%%%%%%%%%%%%%%%%%%%%%%%%%%%%%%%%%%%%%%%%%%%%%%%%%%%%%%%%%%%%%
 
The formula~(\ref{eq:suma}) can be applied in order to perform the
summation in the vertex equation~(\ref{eq:svertex}).  The term associated
with the discontinuities gives (omitting the prefactor $1/2$ and
$\rho(Q)$)
\begin{eqnarray}
	&&\int_{-\infty}^{\infty}\frac{d\xi}{2\pi i}\, n_b(\xi)
	\left\{
	 \frac{G(\xi-i\nu_n,{\mathbf p}+{\mathbf q})}
	   {\xi-q^0-i\omega_m-i\nu_n} \right. \nonumber \\* 
	   &&\quad \times \left[ 
	     G^{\rm adv}(\xi,{\mathbf p}+{\mathbf q})
	     \Lambda_{ij}(\xi-i 0^+,\xi-i\nu_n) 
	    -G^{\rm ret}(\xi,{\mathbf p}+{\mathbf q})
	   \Lambda_{ij}(\xi+i 0^+,\xi-i\nu_n)
	   \right] + \nonumber \\* 
	   && \qquad  \qquad 
	   \frac{G(\xi+i\nu_n,{\mathbf p}+{\mathbf q})}
	   {\xi-q^0-i\omega_m} \nonumber \\* 
	   &&\quad \times \left. \left[ 
	   G^{\rm adv}(\xi,{\mathbf p}+{\mathbf q})
	   \Lambda_{ij}(\xi+i\nu_n,\xi-i0^+) 
	   -G^{\rm ret}(\xi,{\mathbf p}+{\mathbf q})
	   \Lambda_{ij}(\xi+i\nu_n,\xi+i0^+)
	   \right] 
	   \right\},  
\end{eqnarray}
and the pole term associated with the spectral representation of 
the rung  gives
\begin{equation}
n_b(q^0)G(q^0+i\omega_m+i\nu_n,{\mathbf p}+{\mathbf q})
\Lambda_{ij}(q^0+i\omega_m+i\nu_n,q^0+i\omega_m)
G(q^0+i\omega_m,{\mathbf p}+{\mathbf q}),
\end{equation}
where we have omitted the dependence on ${\mathbf p}+{\mathbf q}$ in the 
vertex. 
After this summation, the analytic continuation 
$i\omega_{m}+i\nu_{n}\rightarrow p^0 + \omega +i 0^+, \,
i\omega_{m}\rightarrow p^0-i0^+$ can be explicitly performed 
%in the following way
\begin{eqnarray*}
\Lambda_{ij}(\xi-i 0^+,\xi-i\nu_n)&\rightarrow&
  \Lambda_{ij}(\xi-i 0^+,\xi-\omega-i 0^+), \\ 
\Lambda_{ij}(\xi+i 0^+,\xi-i\nu_n)&\rightarrow&
  \Lambda_{ij}(\xi+i 0^+,\xi-\omega-i 0^+), \\ 
\Lambda_{ij}(\xi+i\nu_n,\xi-i0^+) &\rightarrow& 
 \Lambda_{ij}(\xi+\omega+i 0^+, \xi-i 0^+), \\ 
\Lambda_{ij}(\xi+i\nu_n,\xi+i0^+) &\rightarrow& 
 \Lambda_{ij}(\omega+\xi+i 0^+, \xi+i0^+), \\
 G(\xi-i\nu_n,{\mathbf p}+{\mathbf q}) &\rightarrow& 
   G^{\rm adv}(\xi-\omega,{\mathbf p}+{\mathbf q}), \\	
 G(\xi+i\nu_n,{\mathbf p}+{\mathbf q}) &\rightarrow& 
   G^{\rm ret}(\xi+\omega,{\mathbf p}+{\mathbf q}).  
\end{eqnarray*}
If we neglect the products $G^{\rm adv} G^{\rm adv}$
and $G^{\rm ret} G^{\rm ret}$ as argued before, 
the discontinuity  contribution becomes
\begin{eqnarray}
\int_{-\infty}^{\infty}\frac{d\xi}{2\pi i}&&  
 G^{\rm ret}(\xi+\omega,{\mathbf p}+{\mathbf q})
 \Lambda_{ij}(\xi+\omega+i 0^+, \xi-i0^+)
 G^{\rm adv}(\xi,{\mathbf p}+{\mathbf q})\nonumber \\ 
 &&\times\left[\frac{n_b(\xi)}{\xi-p^0-q^0+i 0^+} - 
         \frac{n_b(\xi+\omega)}{\xi-p^0-q^0-i 0^+}\right], 
\end{eqnarray}
and the required limit $\omega\rightarrow 0$, after the $\xi$ integration, 
reduces to 
\begin{equation}
-n_b(p^0+q^0)G^{\rm ret}(p^0+q^0,{\mathbf p}+{\mathbf q})
 \Lambda_{ij}(p^0+q^0+i 0^+, p^0+q^0-i0^+)
 G^{\rm adv}(p^0+q^0,{\mathbf p}+{\mathbf q}) .
\end{equation}
Adding the pole contribution, one finds the integral 
equation satisfied by the effective vertex
\begin{eqnarray}\label{eq:svertex2}
	{\mathcal D}_{ij}(P) &=& \Lambda_{ij}^{(0)}({\mathbf
	p})+\frac{1}{2} \int \frac{d^4Q}{(2 \pi)^4}\, 
	G^{\rm ret}(P+Q){\mathcal D}_{ij}(P+Q) 
   G^{\rm adv}(P+Q) \nonumber \\
   &&\times 
   \rho(Q)\left[n_b(q^0)-n_b(p^0+q^0) \right] ,
\end{eqnarray}
where we have defined ${\mathcal D}_{ij}(p^0,{\mathbf p}) \equiv
\Lambda_{ij}(p^0 + i 0^+ , p^0 - i 0^+; {\mathbf p})$.

To present a more explicit form of the transport equation, it remains to
analyze the product $G^{\rm ret}(P+Q)G^{\rm adv}(P+Q)$.  
If the momentum $P$ (or $P+Q$) is nearly on-shell, the contribution 
associated with the sum of the crossed products
reduces to a product of two Lorentzians peaked at 
different values or $p^0$. In the limit $\lambda \rightarrow 0$, 
this fact enables us to neglecting  the 
piece $G_+^{\rm ret}G_-^{\rm adv}+
G_-^{\rm ret}G_+^{\rm adv}$, so we may write   
\begin{equation}
G^{\rm ret}(P+Q)G^{\rm adv}(P+Q)= 
\sum_{\pm}\frac{1}{4 |{\mathbf p}+{\mathbf q}|^2}
\frac{1}{(p^0+q^0 \mp |{\mathbf p}+{\mathbf q}|)^2+
\Gamma_{{\mathbf p}+{\mathbf q}}^2}
\end{equation}
which, for $\lambda \rightarrow 0$, behaves as 
\begin{eqnarray}\label{eq:pinch}
G^{\rm ret}(P+Q) G^{\rm adv}(P+Q)&=&
\frac{\pi}{4 |{\mathbf p}+{\mathbf q}|^2\, \Gamma_{{\mathbf p}+{\mathbf q}}}
\left( 
\delta(p^0+q^0-\left|{\mathbf p}+{\mathbf q}\right|)+ 
\delta(p^0+q^0+\left|{\mathbf p}+{\mathbf q}\right|) \right)\nonumber
\\
&=&
\frac{1}{4 |{\mathbf p}+{\mathbf q}| \Gamma_{{\mathbf p}+{\mathbf q}}}\,2 \pi
\,\delta\left( (p^0+q^0)^2-|{\mathbf p} +{\mathbf q}|^2\right)\,.
\end{eqnarray}
Since the imaginary part of the retarded self-energy is odd in the frequency,
the thermal width can be replaced 
by  $-{\rm sgn}(p^0+q^0)\,{\rm Im}\,
\Sigma^{\rm ret}(P+Q)/(2 \left|{\mathbf p}+{\mathbf q}\right|)$, 
and Eq.~(\ref{eq:pinch}) then becomes
\begin{eqnarray}\label{eq:retadv} 
    G^{\rm ret}(P+Q)G^{\rm adv}(P+Q)&=&
-\frac{1}{2\, {\rm Im}\,\Sigma^{\rm ret}(P+Q)}\,2 \pi \,
{\rm sgn}(p^0+q^0) \delta\left( (p^0+q^0)^2-|{\mathbf p} +{\mathbf
q}|^2\right) \nonumber\\
 &=& -\frac{1}{\left(1 + n_b(p^0+q^0)\right)}\frac{1}{2\, {\rm
 Im}\,\Sigma^{\rm ret}(P+Q)}G_0^>(P+Q)\,.
\end{eqnarray}
With the aid of the identity 
\begin{equation}
n_b(q^0)-n_b(p^0+q^0)= n_b(q^0)\left(1+n_b(p^0+q^0)\right)
\left(1-e^{-\beta p^0}\right),
\end{equation}
one arrives to the final form of the transport equation
\begin{equation}\label{eq:svertex3}
	{\mathcal D}_{ij}(P) = 
	  \Lambda_{ij}^{(0)}({\mathbf p})-\left(1-e^{-\beta p^0}\right)
	  \int \frac{d^4Q}{(2 \pi)^4}\,L_{\rm Boltz}(-Q)\,G_0^>(P+Q) 
    \frac{{\mathcal D}_{ij}(P+Q)}{2\, 
 {\rm Im}\,\Sigma^{\rm ret}(P+Q)}\,,
\end{equation}
in complete agreement\footnote{ 
Our notation does not agree
with~\cite{Jeon}, but the conversion is direct: $\Lambda_{i j}^{(0)}
\rightarrow 2 I_{\pi}, {\mathcal D}_{ij}\rightarrow 2 {\mathcal D}_\pi$
and ${\rm Im}\,\Sigma^{\rm ret} \rightarrow -\Sigma_I$.} 
with the results of
Ref.~\cite{Jeon}, 
where the equivalence to the linearized Boltzmann equation has been proven. 
The function ${\mathcal D}_{ij}(P)$ is real, even in 
$p^0$, provided that $\rho(Q)$ is odd in $q^0$. 

The last step in the evaluation of 
shear viscosity is the computation of the integral~(\ref{eq:slope}) for
which only  the on-shell effective vertex ${\mathcal
D}_{ij}(p^0=p,{\mathbf p})$ is required.  The insertion
of~(\ref{eq:retadv}) with $P+Q$ replaced by $P$ into~(\ref{eq:slope}),
and the multiplication by $1/10$ yields the shear viscosity compactly
written as
\begin{equation}
 \eta = -\frac{\beta}{20} \int\frac{d^4P}{(2\pi)^4}\,n_{b}(p^0)
 \Lambda_{i j}^{(0)}({\mathbf p}) G^>(P)
 \frac{{\mathcal D}_{ij}(P)}{2\, {\rm Im}\,\Sigma^{\rm ret}(P)}\,. 
\end{equation}

%%%%%%%%%%%%%%%%%%%%%%%%%%%%%%Softcontributions%%%%%%%%%%%%%%%%%%%%%%%%%%%%%

\section{\label{sec:soft}Soft contributions to the transport equations in gauge
theories}

In the derivation of the transport equations that we try, 
we want emphasize the role played by the
screening effects in the regularization of some infrared divergences
which arise in the small momentum transfer region.
The importance of the dynamical screening in transport phenomena in gauge 
theories 
was first recognized 
in Ref.~\cite{Baym}, where  a linearized collision integral 
free of infrared divergences  
was stated by using only screened interactions mediated by gauge bosons.  
Our aim here is
to derive transport equations by means of the summation of a
restricted set of ladder diagrams, which includes only a specific type
of rungs.  These rungs will consist of appropriate effective bosonic or
fermionic propagators in the hard thermal loop approximation, thus
accounting for the screening of distant collisions between 
all different plasma constituents.  Obviously, this
approximation does not include the effects due to close collisions and
its use requires the imposition of an upper cutoff $q_c$ in the
integrals over the momentum transfer~\cite{Yuan}.  
As a check of consistency, we will
verify that the coefficients of the UV-logarithmic sensitivities to
$q_c$ match to the IR-logarithmic divergences previously computed by 
Arnold, Moore and Yaffe~\cite{Arnold1} in their treatment of the hard
contributions to the collision integrals of the Boltzmann equation.

To elucidate more precisely what type of rungs dominate in the ladder
summation which gives the logarithmic accuracy of transport coefficients, 
it is necessary to estimate the power counting size of the spectral densities
which, in our treatment, are associated with the ladder rungs.  
For instance, consider the spectral density of the rung in the scalar 
theory of the previous section, Eq.~(\ref{eq:bubble}).  
Clearly, it is $O(\lambda^2)$ from the two vertices, and the contribution
$O(\lambda^{-2})$ from the product $G^{\rm ret}G^{\rm adv}$ in
Eq.~(\ref{eq:svertex2}) causes the integral term
in the vertex equation to have a net $g^0$ contribution, as the zero order vertex.

In a gauge theory, there are 
two possible topologies with the same power counting which potentially contribute 
to the leading logarithmic order\footnote{I thank the referee by pointing this out.}. 
They are 
illustrated in Fig.~\ref{fig:rungs}.
The spectral density of the soft gauge
boson exchange ladder, labeled as (a), is $O(1)$ because of  the
$g^2$ suppression from the two vertices, and a $g^{-2} T^{-2}$ 
contribution from the spectral density of the soft gauge boson.  
In the $Q$ integration of
the vertex equation, similar to Eq.~(\ref{eq:svertex2}), 
there is a factor $n_{b}(q^0) = O(g^{-1})$ and a factor $O(g^{-2})$
from the imaginary part of the self-energy in the denominator of the
product $\Delta^{\rm ret}\Delta^{\rm adv}$.  Finally there is a $g^3$
suppression from the soft integration $d^3{\mathbf q}$ (the
contribution of the integration $dq^0$ is cancelled by a delta term
from $\Delta^{\rm ret}\Delta^{\rm adv}$, see
Eqs.~(\ref{eq:fdeltasoft}) and~(\ref{eq:bdeltasoft}) below).  Hence,
the net contribution of the integral term in the vertex equation is
$O(1)$, the same as the zero order vertex,  
when the spectral density of the rung is also $O(1)$.

Similarly, when the horizontal internal lines
are soft gauge boson propagators,  the spectral density of the box type
ladder rung, labeled as (b) in Fig.~\ref{fig:rungs}, is $O(1)$.  
To perform this estimation, it is helpful to write the general 
form of the spectral densities in terms of the squares of the $2\leftrightarrow 2$ 
scattering amplitudes $|\mathcal M|^2$. Such formulas are similar to 
Eq.~(\ref{eq:bubble})
\begin{eqnarray}\label{eq:burbu} 
       \rho_{\mathrm{a}}(P',P) &\propto& \frac{1}{n_b(p'^0-p^0)}
       \int d^4 K d^4 K'
       \delta^{(4)}(P+K-P'-K')|\mathcal M|^2 
       \Delta^>(-K')\Delta^>(K), \\ 
 \label{eq:box}
      \rho_{\mathrm{b}}(K',P) &\propto& \frac{1}{n_b(k'^0-p^0)}
       \int d^4 K d^4 P'
       \delta^{(4)}(P+K-P'-K')|\mathcal M|^2 
       \Delta^>(-P')\Delta^>(K),
\end{eqnarray}
where $P,K,P',K'$ are the on-shell hard momenta for the particle entering
into the scattering processes, and the labels for momenta have been chosen 
with the aim to use the same $|\mathcal M(P,K,P',K')|^2$ within the 
integrations\footnote{ 
We closely follow the notation of Ref.~\cite{Arnold1}.}. For soft momentum transfer $Q=P'-P=K-K'$, this reduces to 
\begin{equation}\label{eq:MM} |
    \mathcal M(P,K,P',K')|^2 \propto g^4  p^2 k^2 |
   {}^\ast\Delta_{\mathrm L}(Q) +
   (\hat{\mathbf{p}}\times\hat{\mathbf{q}})\cdot
   (\hat{\mathbf{k}}\times\hat{\mathbf{q}}) {}^\ast\Delta_{\mathrm
   T}(Q)|^2 \sim O(g^0)\,,
\end{equation}
where ${}^\ast\Delta_{\mathrm{L, T}}$ denotes the longitudinal or
transverse retarded boson gauge propagator.
Note that  $P'$ and $K$ are loop momenta associated with the integrations in 
the vertex equation, and $P$ corresponds to the fixed momentum of the effective vertex. 
In this language, it is easy to recover the power counting size of $\rho_{\mathrm{a}}(P',P)$: 
\begin{eqnarray}
\rho_{\mathrm{a}}(P',P) &\propto& q^0 
 \int_0^\infty dk\, n_{b,f}'(k) \int d\Omega_{\bf k} 
 \delta(q^0-\hat{\mathbf k}\cdot{\mathbf q}) |\mathcal M|^2 
 \propto g^4 p^2 T^2 \frac{q^0}{q} 
 \left[|\Delta_{\mathrm L}(Q)|^2 \right. \nonumber \\
 &&\left.+ \frac{1}{2}\left(1-\frac{q_0^2}{q^2}\right)^2 
      |\Delta_{\mathrm T}(Q)|^2\right] 
 \propto 
 g^2 p^2  \left[\beta_{\mathrm L}(Q)+ 
  \left(1-\frac{q_0^2}{q^2}\right) \beta_{\mathrm T}(Q)\right]
  \sim O(g^0). 
\end{eqnarray}
For the estimation of  $\rho_{\mathrm{b}}(K',P)$, the denominator contributes as 
$g^0$ even though $k'^0$ and $p^0$ are hard energies. The reason for this is that, 
in general, their difference does not correspond to a soft exchange of energy, as shown in 
Fig.~\ref{fig:rungs}(b). Consequently, $\rho_{\mathrm{b}}(K',P) = O(1)$ also.
%%%%%%%%%%%%%%%%%%%%%%%%%%%%%%%%%%%Figure%%%%%%%%%%%%%%%%%%%%%%%%%%%%%%%%%%%
\begin{figure}
    \includegraphics{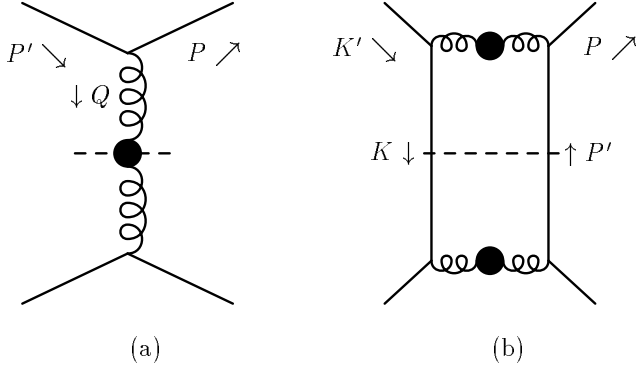} 
    \caption{\label{fig:rungs} Two ladder rung topologies of the same power counting. The
    dashed lines correspond to the cuts giving the spectral densities 
    $\rho_{\mathrm{a}}(P',P)$ and $\rho_{\mathrm{b}}(K',P)$.}
\end{figure}
%%%%%%%%%%%%%%%%%%%%%%%%%%%%%%%%%%%%%%%%%%%%%%%%%%%%%%%%%%%%%%%%%%%%%%%%%%%%
   
However, in
the case of the shear viscosity, the subsequent integration in the
vertex equation over the directions of ${\mathbf k}'$ and the $q^0$ integration 
cancels the contribution of this  box type contribution.
To understand this cancellation, we note that for small momentum
transfer $q \ll p,k$, the energy $\delta$ function which
remains after the three-momentum ${\bf p}'$ integration in Eq.~(\ref{eq:box})
may be expressed as $\delta(q^0-\hat{\mathbf k}'\cdot{\mathbf q})$, and 
\begin{equation}
    \mathcal M(P,K,P',K') \propto g^2 p k  \left[ {}^\ast\Delta_{\mathrm L}(Q) + 
     {}^\ast\Delta_{\mathrm T}(Q)\, \left(1-\frac{q_0^2}{q^2} \right) \cos
     \phi \right]\,,
   \end{equation}
where $\phi$ is the  angle  
between the planes $({\mathbf p},{\mathbf q})$ 
and $({\mathbf k}',{\mathbf q})$. 
On the other hand, the angular dependence of $\Lambda_{\pi}(K')$
within the integration over $K'$ in the vertex equation is
given by the contraction of $\Lambda_{\pi}(K')$ with $\Lambda_{\pi}(P)$,
which is proportional to $P_{2}(\cos\theta_{pk'})$, with $P_{2}$ 
the second Legendre polynomial. For small $q$, 
the required cosine becomes~\cite{Arnold1}
\begin{equation}
     \cos \theta_{pk'} \simeq \frac{q_0^2}{q^2} +
     \left(1-\frac{q_0^2}{q^2} \right) \cos \phi\,.
\end{equation}
Without screening
effects, the boson propagators reduce to 
${}^\ast\Delta_{L}(Q) =-1/q^2$ and ${}^\ast\Delta_{T}(Q) = -1/(q_0^2-q^2)$.  
Thus, the angular integration 
over the azimuthal angle of ${\mathbf k}'$ followed by the $k^0$ (or $q^0$) 
integration
turns out to be proportional to
\begin{equation}
\int_{-q}^q dq^0 \int_0^{2 \pi} d\phi\, P_{2}(\cos\theta_{pk'}) (1-\cos
\phi)^2 = 0\,.
\end{equation}
Here, the integration over the spatial range $-q<q^0< q$ is 
enforced by the $\delta(q^0 -\hat{\mathbf k}'\cdot{\mathbf q})$.
This means that the unscreened box rungs do not contribute to the
leading logarithmic order and, consequently, 
box ladder rungs made of two soft gauge
boson propagators are irrelevant in order to compute the 
soft contribution to the shear viscosity.

For the electrical conductivity at zero chemical potentials, the
cancellation of these box ladder rungs comes from the charge-conjugation
invariance.  For each fixed $P$, we have two contributions to the
integral term of the vertex equation, corresponding to the insertions of
fermions and antifermions through the loop. These contributions are of
opposite sign, so their sum vanishes. 
   
Similarly, there are other box ladder rungs in which the horizontal lines are soft
fermion propagators, which are comparable to 
the soft fermion exchange ladders. They do not contribute either to the leading
logarithmic order  because of the cancellation under angular integration
of their unscreened counterparts.  Now, the unscreened squared
amplitudes are 
$|\mathcal M|^2 \propto (1-\cos \phi)/q^2$~\cite{Arnold1}, and the relevant 
integration is 
\begin{equation}
\int_{-q}^q dq^0 \int_0^{2 \pi} d\phi\, P_{l}(\cos\theta_{pk'}) (1-\cos
\phi) = 0\,,
\end{equation}
with $l = 1, 2$ for the electrical conductivity and the shear viscosity, respectively.  

Next, let us consider the equations for the effective 
vertex $\Lambda_{A}$ of 
a given fermionic species and 
the effective vertex $\Lambda_{A}^{j m}$ for a given boson gauge.  
Here, the 
superindices $(jm)$ denote spatial indices corresponding to the 
boson propagators in Coulomb gauge to be joined to the vertex, 
and  $A$ collectively denotes the indices corresponding to the 
insertion of $j_i$ or $\pi_{ik}$.
We  still does not 
explicitly indicate any spin, color or flavor indices.  
The equations for 
the effective vertices are illustrated in Figs.~\ref{fig:vf}
and~\ref{fig:vb}.  These equations sum all ladders consisting of
rungs made of a HTL gauge boson propagator, and also a HTL fermion
propagator.  The exchange of a soft gauge boson takes place in
scattering between fermions and gauge bosons in non-abelian theories,
while the exchange of a soft fermion enters in an annihilation process
into a fermionic pair, and the inverse process of creation. 
%%%%%%%%%%%%%%%%%%%%%%%%%%%%%%%%%%%Figure%%%%%%%%%%%%%%%%%%%%%%%%%%%%%%%%%%%
\begin{figure}
    \includegraphics{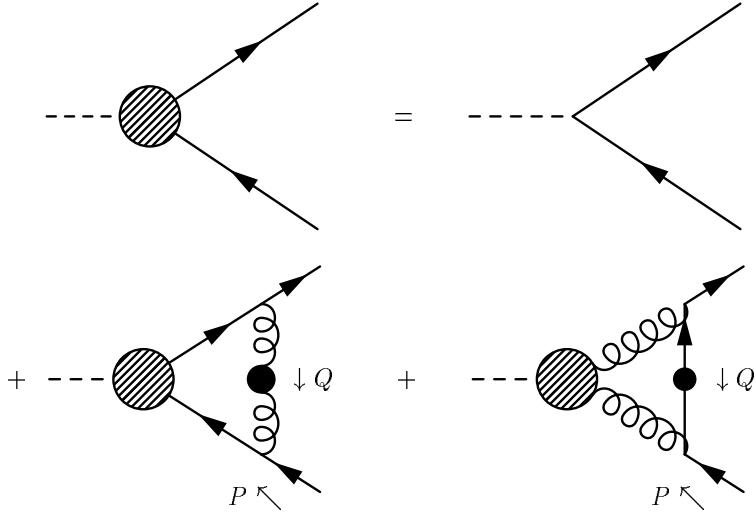} 
    \caption{\label{fig:vf}Equation for the soft contribution to the 
     fermionic effective vertex. For the case of the electrical conductivity
     to leading order in $e$, 
     the soft boson exchanged corresponds to a photon 
     and the last graph is zero.}
\end{figure}
%%%%%%%%%%%%%%%%%%%%%%%%%%%%%%%%%%%%%%%%%%%%%%%%%%%%%%%%%%%%%%%%%%%%%%%%%%%%
%%%%%%%%%%%%%%%%%%%%%%%%%%%%%%%%%%%Figure%%%%%%%%%%%%%%%%%%%%%%%%%%%%%%%%%%%
\begin{figure}
    \includegraphics{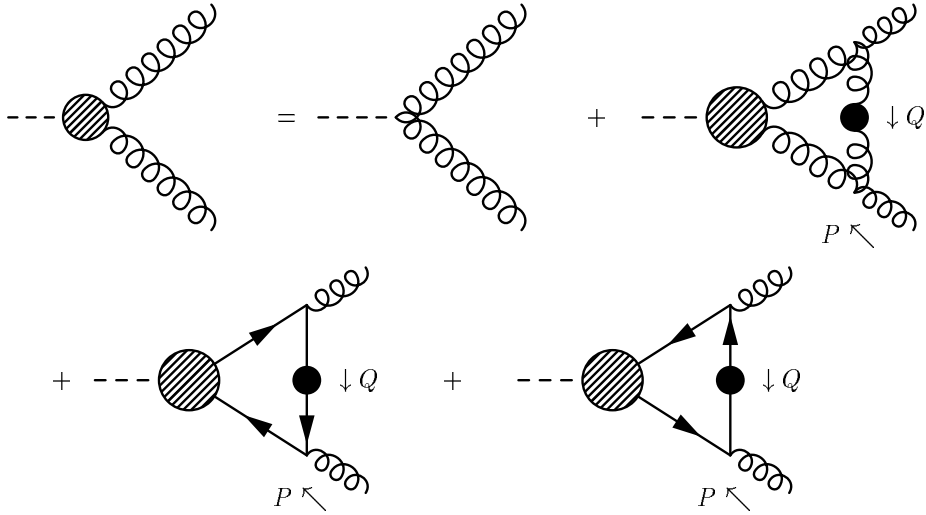} 
    \caption{\label{fig:vb}Equation for the soft contribution to the 
    bosonic effective vertex. In the case of the electrical conductivity, 
    all graphs are null.}
\end{figure}
%%%%%%%%%%%%%%%%%%%%%%%%%%%%%%%%%%%%%%%%%%%%%%%%%%%%%%%%%%%%%%%%%%%%%%%%%%%%

By following a completely similar treatment 
to that we have presented for the scalar theory 
for the summation and analytical continuation to real frequencies, 
one arrives to the coupled equations for the 
effective vertices 
\begin{eqnarray}\label{eq:fvertex2}
	\Lambda_{A}(P) = 
	 \Lambda_{A}^{(0)}({\mathbf p})&+&  
	 g^2 \int\frac{d^4 Q}{(2 \pi)^4}\, \gamma^\mu S^{\rm
	 ret}(P+Q)\Lambda_{A}(P+Q) S^{\rm adv}(P+Q) \gamma^\nu \nonumber \\* &&
	 \times {}^\ast \rho_{\mu \nu}(Q) \left[n_b(q^0)+n_f(p^0+q^0)\right]
	 \nonumber\\* 
	 &+& g^2 \int \frac{d^4 Q}{(2 \pi)^4}\, G_{ij}^{\rm ret}(P+Q)
	 \Lambda_{A}^{j m}(P+Q) G_{m n}^{\rm adv}(P+Q) \gamma^i \nonumber \\*
	 && \times {}^\ast \rho(\omega,{\mathbf q}) \gamma^n
	 \left[n_f(q^0)+n_b(p^0+q^0)\right],
\end{eqnarray}
and 
\begin{eqnarray}\label{eq:bvertex2}
\Lambda_{A}^{jm}(P) &=& 
	 \Lambda_{A}^{(0)\, jm}({\mathbf p})+  
	 \int\frac{d^4 Q}{(2 \pi)^4}\, 
	 V^{j \mu k}(P,Q,-P-Q)V^{m \nu n}(-P,-Q,P+Q)\nonumber \\* 
	 &&\times G_{ki}^{\rm ret}(P+Q)
	 \Lambda_{A}^{il}(P+Q) 
	 G_{ln}^{\rm adv}(P+Q) {}^\ast \rho_{\mu \nu}(Q)
   \left[n_b(q^0)-n_b(p^0+q^0)\right] \nonumber \\*
   &&+ g^2 \int\frac{d^4 Q}{(2 \pi)^4}\, {\rm tr}\left\{ 
	 \gamma^j S^{\rm ret}(P+Q)\Lambda_{A}(P+Q) 
	 S^{\rm adv}(P+Q) \gamma^m  {}^\ast \rho(Q) \right.  \\*
	 && -\left.  \gamma^m S^{\rm ret}(-P-Q)\Lambda_{A}(-P-Q) S^{\rm
	 adv}(-P-Q) \gamma^j {}^\ast \rho(Q)\right\} \nonumber
	 \left[n_f(q^0)-n_f(p^0+q^0)\right] .
 \end{eqnarray} 
Here ${}^\ast \rho_{\mu \nu}(Q)$ and ${}^\ast \rho(Q)$ denote the
spectral densities of the soft gauge boson and the fermion and, as usual, 
$\Lambda_{A}(P)\equiv \Lambda_{A}(p^0+i0^+,p^0-i0^+;{\mathbf p})$.  
Note that we have used the parity properties ${}^\ast \rho(Q)={}^\ast \rho(-Q)$ 
and $n_f(-q^0)-n_f(-p^0-q^0)=-\left[n_f(q^0)-n_f(p^0+q^0)\right]$. 
These equations are entirely similar to the transport equation for the
scalar theory.  As there, the occupation numbers $n_{b,f}(p^0+q^0)$
arise from the branch cut contributions, and the identities
$n_{b,f}(\xi+n \pi i)=-n_{f,b}(\xi)$ with $n$ odd have been used, if
necessary.  Further progress requires to examine in detail the
explicit structure of the products $S^{\rm ret}\Lambda_{A} S^{\rm
adv}$ and $G^{\rm ret}\Lambda_{A} G^{\rm adv}$.

The substitution of the two pieces of the Green's functions in the product
    $S^{\rm{ret}}S^{\rm{adv}}$ or $G^{\rm{ret}}G^{\rm{adv}}$
    yields four terms, two of them 
    $\Delta_{\pm}^{\rm{ret}}\Delta_{\mp}^{\rm{adv}}$ or 
    $G_{\pm}^{\rm{ret}}G_{\mp}^{\rm{adv}}$ 
    can be directly dropped in the limit
    $g\rightarrow0$.  Now, by computing the soft contribution to the
    transport properties, we have only to consider one of the two terms, $++$
    or $--$, depending on whether the external momentum corresponds to 
    $p^0=p$ or  
    $p^0=-p$.  This important simplification is due to the fact that, when
    the external momentum $P$ is on-shell, let's say $p^0=p$, the other
    sheet of the mass-shell can not be connected by a soft momentum
    transfer.  Clearly, a hard momentum $Q$ is required in order to
    create two propagating on-shell particles both with hard momentum.  In
    contrast, for the case of the scalar theory, we have retained both
    terms in Eq.~(\ref{eq:pinch}) because there, we have not made
    distinction between soft and hard momentum transfers.  Thus, when $P$
    is on-shell and $Q\ll P$, we can make the approximations
   \begin{equation}\label{eq:fdeltasoft}
      \Delta_{\pm}^{\rm{ret}}(P+Q)\Delta_{\pm}^{\rm{adv}}(P+Q) =
      \frac{\pi}{\gamma_{{\mathbf p}+{\mathbf q}}}\,
      \delta(p^0+q^0\mp|{\mathbf{p}}+{\mathbf{q}}|) \simeq
      \frac{\pi}{\gamma_{{\mathbf p}+{\mathbf q}}}\, \delta(q^0 \mp
      \hat{\mathbf{p}}\cdot{\mathbf{q}}),
   \end{equation}
   and
   \begin{equation}\label{eq:bdeltasoft}
      G_{\pm}^{\rm{ret}}(P+Q) G_{\pm}^{\rm{adv}}(P+Q) 
      %= \frac{\pi}{4 p^2\,\Gamma_{{\mathbf p}+{\mathbf q}}}\,
      %\delta(p^0+q^0\mp|{\mathbf{p}}+{\mathbf{q}}|) 
      \simeq
      \frac{\pi}{4 p\, |{\mathbf p}+{\mathbf q}|\,\Gamma_{{\mathbf p}+{\mathbf q}}}\, 
	  \delta(q^0
      \mp \hat{\mathbf{p}}\cdot{\mathbf{q}}).
   \end{equation}
 
 On the other hand, in the high temperature limit when all masses are
 negligible, the theory is chirally invariant.  As a consequence, the
 projection operators $h_{\pm}({\hat{\mathbf p}})$ can be expressed in
 terms of simultaneous eigenspinors of chirality and helicity,
 \begin{eqnarray}\label{eq:hplus}
	h_{+}({\hat{\mathbf p}})&=& 
	  u({\hat{\mathbf p}},+) \overline{u}({\hat{\mathbf p}},+)+
	  u({\hat{\mathbf p}},-) \overline{u}({\hat{\mathbf p}},-)\, , \\ 
	\label{eq:hminus}
	h_{-}({\hat{\mathbf p}})&=& 
	  v({\hat{\mathbf p}},+) \overline{v}({\hat{\mathbf p}},+)+
	  v({\hat{\mathbf p}},-) \overline{v}({\hat{\mathbf p}},-)\, ,
 \end{eqnarray}
 where the $u\,(v)$-spinors have the chirality equal (opposite) to the 
 helicity. 
For example, in the chiral representation, 
%with spin up ($+$) or down ($-$) 
%and the momentum along the $3$-axis 
with the momentum along the $3$-axis
\begin{equation}
u({\hat{\mathbf p}},+)=\left(\begin{array}{c}
                       0\\0\\1\\0
                       \end{array} \right),\quad
u({\hat{\mathbf p}},-)=\left(\begin{array}{c}
                       0\\1\\0\\0
                       \end{array} \right)\quad 
v({\hat{\mathbf p}},+)=\left(\begin{array}{c}
                       1\\0\\0\\0
                       \end{array} \right),\quad
v({\hat{\mathbf p}},-)=\left(\begin{array}{c}
                       0\\0\\0\\1
                       \end{array} \right).                     
\end{equation}

In the case of the electrical conductivity, the zero-order effective
vertices are
\begin{eqnarray}\label{eq:zerofcond}
\Lambda_{i}^{(0),\,s}&=& q_s e \gamma^i , \\ 
\Lambda_{i}^{(0) j m}&=& 0,
\end{eqnarray}
where $q_s$ is the charge of the fermionic constituent $s$ in units of $e$. 
For the shear viscosity, the insertion at zero momentum of a 
spatially transverse energy-momentum tensor yields the zero-order vertices 
\begin{eqnarray}\label{eq:zerofsh}
\Lambda_{i j}^{(0)}({\mathbf p}) &=& \frac{1}{2}\left(\gamma^i p^j+ \gamma^j
p^i - \frac{2}{3}\, {\bm{\gamma}}\cdot{\mathbf p}
\delta^{i j}\right), \\
\label{eq:zerobsh}
\Lambda_{i k}^{(0) j m}({\mathbf p})&=& 2 \left(p^i p^k - \frac{p^2}{3}
\delta^{ik} \right) \delta^{jm} .
\end{eqnarray}
Both of the fermionic vertices are linear in the $\gamma$-matrices, 
and this linear dependence 
is preserved by summing ladder diagrams because each added rung does not 
introduce any extra dependence. 
Thus, the fermionic effective vertices, which appear sandwiched between  
the projection operators $h_{\pm}({\hat{\mathbf p}})$ and 
$h_{\pm}(\widehat{{\mathbf p}+{\mathbf q}})$  require to consider a 
$\gamma$-matrix  between 
eigenspinors of all possible chiralities and helicities. 
Obviously, the combinations between spinors of different chirality, 
as $\overline{u}(\pm)\gamma^\mu u(\mp)$ or 
$\overline{v}(\pm)\gamma^\mu u(\pm)$, are zero.
Moreover, %as argued before, 
the combinations mixing the particle and antiparticle
mass shells,  as $\overline{v}(\widehat{{\mathbf p}+{\mathbf
q}},\pm)\gamma^\mu u({\hat{\mathbf p}},\mp)$ or  
$\overline{u}(\widehat{{\mathbf p}+{\mathbf
q}},\pm)\gamma^\mu v({\hat{\mathbf p}},\mp)$,  do not need to be retained
since they can not be connected by a soft momentum transfer.
Therefore, we are left with the chirality-independent combinations
\begin{eqnarray}\label{eq:gordon1}
   &&\overline{u}(\widehat{{\mathbf p}+{\mathbf q}},\pm)\gamma^0 
     u({\hat{\mathbf p}},\pm) = 1 + O(q/p), \qquad 
   \overline{u}(\widehat{{\mathbf p}+{\mathbf q}},\pm)\gamma^i 
     u({\hat{\mathbf p}},\pm) = \hat{p}^i + O(q/p),   \\ 
\label{eq:gordon2}
   &&\overline{v}(\widehat{{\mathbf p}+{\mathbf q}},\pm)\gamma^0 
     v({\hat{\mathbf p}},\pm) = 1 + O(q/p), \qquad 
   \overline{v}(\widehat{{\mathbf p}+{\mathbf q}},\pm)\gamma^i 
     v({\hat{\mathbf p}},\pm) = -\hat{p}^i + O(q/p),
\end{eqnarray}
corresponding to the lepton (quark) and the anti-lepton (anti-quark), respectively.

\subsection{Soft contribution to the transport equation for the
electrical conductivity}
Now, we are ready to write more explicitly the 
transport equations~(\ref{eq:fvertex2}) and~(\ref{eq:bvertex2}).  For the
case of the electrical conductivity, it is suggested to define the
non-amputated on-shell vertices for a given charged fermionic species
$s$
\begin{eqnarray}\label{eq:part}
   {\mathcal D}_i^{s^-}({\mathbf p}) &\equiv& 
   \overline{u}(\hat{{\mathbf p}},\pm) 
   \Lambda_{i}^s(p^0=p,{\mathbf p}) 
   u({\hat{\mathbf p}},\pm) ,  \\
   \label{anti}
    {\mathcal D}_i^{s^+}({\mathbf p}) &\equiv& 
	\overline{v}(\hat{{\mathbf p}},\pm)
   \Lambda_{i}^s(p^0=-p,-{\mathbf p}) 
   v({\hat{\mathbf p}},\pm), 
\end{eqnarray}
with the corresponding zero-order vertices 
which follow from Eqs.~(\ref{eq:gordon1}) and~(\ref{eq:gordon2}).  
Here, the factor $g^2$ in the front of the integral in Eq.~(\ref{eq:fvertex2}) 
must be replaced by
$q_{s}^2 e^2$.  At zero electrical charge, the bosonic effective
vertex for electrical conductivity does not enter because its
zero-order vanishes, and Furry's theorem ensures the vanishing of the
term within the trace in Eq.~(\ref{eq:bvertex2}).  This leaves a
single decoupled equation for the fermion vertex.

Next, we fix the external frequency $p^0=p$. After 
multiply both sides of  the equation~(\ref{eq:fvertex2}) by the $u$-eigenspinors, 
and expand the integrand using~(\ref{eq:hplus})and~(\ref{eq:fdeltasoft})
with $++$, we obtain a term proportional to
\begin{eqnarray}
\delta(q^0&-&\hat{{\mathbf p}}\cdot{\mathbf q})
\overline{u}(\hat{{\mathbf p}},\pm)\gamma^\mu 
u(\widehat{{\mathbf p}+{\mathbf q}},\pm)\, 
\overline{u}(\widehat{{\mathbf p}+{\mathbf q}},\pm)\gamma^\nu
u(\hat{{\mathbf p}},\pm)\,{}^\ast \rho_{\mu\nu}(Q) = \nonumber \\
&&\delta(q^0-\hat{{\mathbf p}}\cdot{\mathbf q}) 
\left[ \beta_{\rm L}(Q) + 
\left( 1- \frac{q_0^2}{q^2}\right)\beta_{\rm T}(Q)\right]+O[q/p],
\end{eqnarray} 
where 
the delta function has enforced a space-like momentum transfer momentum corresponding 
to Landau damping, and we have used the approximations~(\ref{eq:gordon1}). 
Finally, the substitution 
$\Lambda^{i}(\left|{\mathbf p}\right|+\hat{{\mathbf p}}\cdot{\mathbf q},
{\mathbf p}+{\mathbf q})=
\Lambda^{i}(\left|{\mathbf p}+{\mathbf q}\right|,{\mathbf p}+{\mathbf q})$, 
valid at soft momentum 
%and 
%\begin{equation}
%n_b(q^0)+n_f(p^0+q^0)=\frac{T}{q^0} -\frac{1}{2}+n_f(p)+O(q^0/p^0),  
%\end{equation} 
gives the equation
\begin{eqnarray}\label{eq:transport1}
{\mathcal D}_i^{s^-}({\mathbf p})&=&q_s e\,  \hat{p}^i+ 
q_s^2 e^2 \int\frac{d^4Q}{(2\pi)^4}\,
\pi \delta(q^0-\hat{{\mathbf p}}\cdot{\mathbf q}) 
\left(\frac{T}{q^0} -\frac{1}{2}+n_f(p)\right) \nonumber \\  
&&\times \left[\beta_{\rm L}(Q) + 
\left( 1- \frac{q_0^2}{q^2}\right)\beta_{\rm T}(Q) \right]
\frac{{\mathcal D}_i^{s^-}({\mathbf p}+{\mathbf q})}
{\gamma_{{\mathbf p}+{\mathbf q}}^s},
\end{eqnarray}
where we have neglected the $O(q^0/p)$ terms  in the occupation numbers. 
Using the parity properties of the zero-order vertex and noting that 
the spectral density ${}^\ast\rho_{\mu\nu}(Q)$ is odd in $q^0$, 
one may easily show from~(\ref{eq:bvertex2}) that 
${\mathcal D}_i^{s^-}({\mathbf p})= -{\mathcal D}_i^{s^+}({\mathbf p})$.
This equation 
is formally similar to the integral equation for the complete 
leading order of the photon emission rate from the quark-gluon plasma,  
which has been derived in Ref.~\cite{Arnold2}.

Following a similar treatment to that used by Arnold, Moore and 
Yaffe~\cite{Arnold2}, we may reduce a bit more the transport equation. 
This proceeds by insertion of the integral for the thermal width of a
hard fermion.  As we have already said, the thermal width receives the
leading order contribution from two-body scattering between fermions
by exchange of a soft photon.  For a given fermionic species, it is
given by the integral~\cite{Blaizot}
 \begin{equation}\label{eq:damping1}
 \gamma_{{\mathbf p}}^{(2),s} = q_{s}^2 e^2 \int\frac{d^4Q}{(2\pi)^4}\,
 \pi \delta(q^0-\hat{{\mathbf p}}\cdot{\mathbf q})\frac{T}{q^0}
 \left[\beta_{\rm L}(Q)+
\left(1- \frac{q_0^2}{q^2}\right) \beta_{\rm T}(Q)\right] . 
 \end{equation}
 Note that the value of this integral remains unaltered when the 
 term $T/q^0$ of the integrand is replaced  by  $T/q^0-1/2 + n_f(p)$. 
 This is due 
 to the fact that the added term $-1/2 + n_f(p)$ is here multiplied by an odd function 
 of $q^0$ and, consequently, is irrelevant. However, this is not true for 
 the corresponding term in~(\ref{eq:transport1}), since the vectorial 
 dependence of ${\mathcal D}_i$ on ${\mathbf p}+{\mathbf q}$ can give 
 odd contributions in ${\mathbf p}\cdot{\mathbf q}$.
   
 Besides the dominant contribution, the thermal width receives another 
 contribution coming from 
 two-body conversion processes which  
 also give rise to a leading-log term in the
 transport coefficients.  This contribution to the thermal width
 corresponds to the imaginary part of the diagram of Fig.~\ref{fig:selffermi}, 
 and
 it is expressed by the integral
 \begin{eqnarray}\label{eq:damping2}
 \gamma_{{\mathbf p}}^{(4),s} &=& q_{s}^2 e^2 
\left(\frac{1}{2}+n_{b}(\left|{\mathbf p}\right|) \right) 
\frac{\pi}{2p}
\int\frac{d^4Q}{(2\pi)^4}\,\delta(q^0-\hat{{\mathbf p}}\cdot{\mathbf
q}) \nonumber \\*
&&\times \left[ \beta_+(Q)\left(1-\frac{q^0}{q}\right) 
              + \beta_-(Q)\left(1+\frac{q^0}{q}\right)\right] \nonumber \\*
&=&\frac{q_s^2 e^2 \omega_s^2}{8\pi p}
\left(1+2n_b(\left|{\mathbf p}\right|)\right)\left[
\ln\left(\frac{q_c}{\sqrt{2} \omega_s}\right)-1+\ln 2\right],
\end{eqnarray} 
where $q_c$ is the upper limit of integration which separates 
semihard and hard momentum transfers, and
$\omega_s^2=q_s^2 e^2 T^2/8$ is the plasma frequency for 
the fermionic species $s$. 
%%%%%%%%%%%%%%%%%%%%%%%%%%%%%%%%%%%Figure%%%%%%%%%%%%%%%%%%%%%%%%%%%%%%%%%%%
\begin{figure}
    \includegraphics{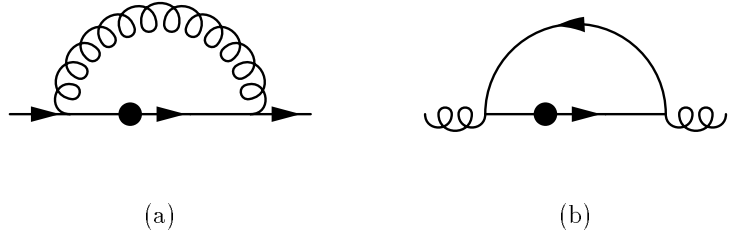} 
    \caption{\label{fig:selffermi}Subleading contributions to the self-energy of 
     a hard quark~(a) and  a gluon~(b).}
\end{figure}
%%%%%%%%%%%%%%%%%%%%%%%%%%%%%%%%%%%%%%%%%%%%%%%%%%%%%%%%%%%%%%%%%%%%%%%%%%%%

By inserting the expression~(\ref{eq:damping1}) 
into~(\ref{eq:transport1}), after defining the quantities
\begin{equation}\label{eq:chi0}
      \chi_i^{s^{-,+}}({\mathbf p}) \equiv 
      %q_s e \frac{{\mathcal D}_i^{s^{-,+}}({\mathbf p})}{\gamma_{\mathbf p}^s} ,  
       \frac{q_s e}{\gamma_{\mathbf p}^s}\, {\mathcal D}_i^{s^{-,+}}({\mathbf p}) ,       
\end{equation}
one obtains a more convenient expression  
\begin{eqnarray}\label{eq:chi}
q_s^2 e^2\hat{p}^i&=& 
\gamma_{{\mathbf p}}^{(4),s}\, \chi_i^{s^{-}}({\mathbf p}) + 
q_s^2 e^2 \int\frac{d^4Q}{(2\pi)^4}\,
\pi \delta(q^0-\hat{{\mathbf p}}\cdot{\mathbf q})
\left(\frac{T}{q^0} -\frac{1}{2}+n_f(p)\right) \nonumber \\*
&&\times\left[\beta_{\rm L}(Q)+
\left(1- \frac{q_0^2}{{\mathbf q}^2}\right)\beta_{\rm T}(Q) \right]
\left[\chi_i^{s^{-}}({\mathbf p})-
  \chi_i^{s^{-}}({\mathbf p}+{\mathbf q})   \right], 
\end{eqnarray}
for the soft contribution to the transport equation in the 
case of electrical conductivity. 

An important point to notice here concerns to the absence 
of some subleading corrections to the hard particle width,
whose size  is comparable to that included in Eq.~(\ref{eq:damping2}).
Such  corrections  of $O(g^3)$ and  $O(g^4 \ln g^{-1})$
arise from the subleading part of 
the bosonic occupation number $n_b(q^0)$ in the integrand 
of Eq.~(\ref{eq:damping1}), and they are associated with  
\begin{equation}\label{eq:ML}
        n_b(q^0) - \frac{T}{q^0} + \frac{1}{2}=
        2 T \sum_{n=1}^{\infty} \frac{q^0}{q_0^2 + (2 \pi n T)^2}.
\end{equation}
To understand this absence, consider the transport 
equation~(\ref{eq:transport1}) when $\mathcal D/\gamma$ has been replaced 
by $\chi$, as in Eq.~(\ref{eq:chi0}), and the kernel 
still contains 
$n_b(q^0)+n_f(p)$.
Now, if in the product $\gamma_{\bf p} \chi({\bf p})$ one were to include 
the subleading terms from Eq.~(\ref{eq:ML}), 
the same terms would have to be included 
into the occupation number in the kernel of the integral.
For small $q$, there is a cancellation between them, and 
the resulting transport equation
would be the same as that in Eq.~(\ref{eq:chi}) with 
$T/q^0-1/2+n_f(p)$ replaced by $n_b(q^0)+n_f(p)$. 
However, when the momentum  $p$ is hard, $n_f(p) = O(1)$, and 
other pole terms different from zero in the occupation number are 
$O(g)$. Hence, they may be ignored.

\subsection{Soft contribution to the transport equations for shear viscosity}

Next, we consider the shear viscosity in a $SU(N_c)$ 
gauge theory with $N_f$ fermion fields in 
a given irreducible representation $r$ of dimension $d(r)$. 
The generator matrices are 
denoted by $t_r^a$, and the normalization for this representation is defined 
by the constant $C(r)$ in ${\rm tr}(t_r^a t_r^b)= C(r) \delta^{a b}$. 
The quadratic Casimir operator is denoted by $C_2(r)$.  

Since the gluon effective vertex 
$\Lambda_{ik}^{j m, ab}$ is always joined to a pair of  
transverse projectors of the gauge propagators, it is  
useful to define the non-amputated on-shell gluon vertex 
${\mathcal D}_{ik}^g$ by 
\begin{equation}
P_T^{rj}(\hat{\mathbf p})
\Lambda_{ik}^{j m, ab}(p^0=\left|{\mathbf p}\right|,{\mathbf p})
P_T^{m s}(\hat{\mathbf p})
\equiv {\mathcal D}_{ik}^g({\mathbf p}) 
P_{\rm T}^{r s}(\hat{\mathbf p})
\delta^{a b}, 
\end{equation}
and the zero-order vertex corresponding to Eq.~(\ref{eq:zerobsh}), 
${\mathcal D}_{ik}^{(0)\,g}({\mathbf p})= 2 (p^i p^k - \delta^{ik} 
{\mathbf p}^2/3)$. On the other hand, and as before, we define 
the the non-amputated on-shell quark vertex 
${\mathcal D}_{ik}^q$ by
\begin{equation}
{\mathcal D}_{ik}^q({\mathbf p})\equiv \overline{u}(\hat{\mathbf p},\pm)
  \Lambda_{ik}(p^0=p,{\mathbf p}) 
  u(\hat{\mathbf p},\pm),
\end{equation}
and the zero-order vertex corresponding to Eq.~(\ref{eq:zerofsh}),
${\mathcal D}_{ik}^{(0)\,q}({\mathbf p})= |{\mathbf p}|(\hat{p}^i
\hat{p}^k - \delta^{ik}/3)$.  We note that the similar vertex for
anti-quarks, ${\mathcal D}^{\overline{q}}= \overline{v}\Lambda(-P)v$,
turns out to be the same as for quarks, as it is easily checked by
examining the corresponding zero-order vertices, and noting that each
added rung in the iteration does not break this property.

Now, in order to derive the transport equations for 
${\mathcal D}_{ik}^{q,\,g}$, it only remains to perform the 
appropriate contraction of Eqs.~(\ref{eq:fvertex2}) and (\ref{eq:bvertex2})
with a pair of $u$-spinors and a pair of transverse projectors,
respectively.  Using the approximations
\begin{eqnarray}
&&\overline{u}(\hat{\mathbf p},\pm)\gamma^m \left[
\beta_+(Q) h_+({\mathbf q}) +
\beta_-(Q) h_-({\mathbf q})\right]
\gamma^n u(\hat{\mathbf p},\pm)
P_{\rm T}^{m n}(\widehat{{\mathbf p}+{\mathbf q}})= \nonumber \\
&& \qquad \qquad \beta_+(Q)\left(1-\hat{\mathbf p}\cdot\hat{\mathbf q}\right)+
   \beta_-(Q)\left(1+\hat{\mathbf p}\cdot\hat{\mathbf q}\right)
   + O(q/p), \\
&&{\rm tr}\left\{
\gamma^m h_+(\widehat{{\mathbf p}+{\mathbf q}}) \gamma^n
 \left[\beta_+(Q) h_+({\mathbf q}) +
       \beta_-(Q) h_-({\mathbf q}) \right] \right\}
       P_{\rm T}^{m r}(\hat{\mathbf p}) 
       P_{\rm T}^{n s}(\hat{\mathbf p})= \nonumber \\
&& \qquad \qquad \left[\beta_+(Q)\left(1-\hat{\mathbf p}\cdot\hat{\mathbf q}\right)+
   \beta_-(Q)\left(1+\hat{\mathbf p}\cdot\hat{\mathbf q}\right)\right]
   P_{\rm T}^{r s}(\hat{\mathbf p})+ O(q/p),
\end{eqnarray}
%valid when $\left|{\mathbf q} \right \ll \left||{\mathbf p}\right$ 
and
\begin{eqnarray} 
    &&f^{acd}f^{bcd} V^{j\mu k}(P,Q,-P-Q)
     V^{m \nu n}(-P,-Q,P+Q) {}^\ast\rho_{\mu \nu}(Q) 
     P_{\rm T}^{k n}(\widehat{{\mathbf p}+{\mathbf q}}) 
     P_{\rm T}^{j r}(\hat{\mathbf p}) 
     P_{\rm T}^{m s}(\hat{\mathbf p}) = \nonumber \\
&& \qquad \qquad 4 g^2 N_c\,{\mathbf p}^2 \left[ \beta_{\rm L}(Q)+
\left(1- \hat{\mathbf p}\cdot\hat{\mathbf q}^2\right) \beta_{\rm
T}(Q)\right] P_{\rm T}^{r s}(\hat{\mathbf p}) \delta^{ab}+ O(q/p),
\end{eqnarray}
one finds 
\begin{eqnarray}\label{eq:tquark}
{\mathcal D}_{ik}^q({\mathbf p})&=&{\mathcal D}_{ik}^{(0)\,q}({\mathbf p})
+ g^2 C_2(r) \int\frac{d^4Q}{(2\pi)^4}\,
\pi \delta(q^0-\hat{{\mathbf p}}\cdot{\mathbf q})\,
{\mathcal B}(Q) 
\left(\frac{T}{q^0}-\frac{1}{2}+n_f(p) \right)
\frac{{\mathcal D}_{ik}^q ({\mathbf p}+{\mathbf q})}
{\gamma_{{\mathbf p}+{\mathbf q}}}\nonumber  \\ 
&&+g^2 C_2(r) \int\frac{d^4Q}{(2\pi)^4}\,
\pi \delta(q^0-\hat{{\mathbf p}}\cdot{\mathbf q})\,
{\mathcal F}(Q)
 \left(\frac{1}{2}+n_b(p)\right) 
 \frac{{\mathcal D}_{ik}^g ({\mathbf p}+{\mathbf q})}
{4 p\,\left|{\mathbf p}+{\mathbf q}\right| 
   \Gamma_{{\mathbf p}+{\mathbf q}}},          
\end{eqnarray}
and a completely similar for the gluon vertex 
\begin{eqnarray}\label{eq:tgluon}
    {\mathcal D}_{ik}^g({\mathbf p})&=&{\mathcal
D}_{ik}^{(0)\,g}({\mathbf p}) + g^2 N_c \left|{\mathbf p}\right|
\int\frac{d^4Q}{(2\pi)^4}\, \pi \delta(q^0-\hat{{\mathbf
p}}\cdot{\mathbf q})\, {\mathcal B}(Q) 
\left(\frac{T}{q^0}-\frac{1}{2}-n_b(p) \right)
\frac{{\mathcal D}_{ik}^g ({\mathbf p}+{\mathbf q})}
{\left|{\mathbf p}+{\mathbf q}\right|\Gamma_{{\mathbf p}+{\mathbf q}}}\nonumber  \\ 
&&+2 g^2 N_f C(r) 
\int\frac{d^4Q}{(2\pi)^4}\,
\pi \delta(q^0-\hat{{\mathbf p}}\cdot{\mathbf q})\,
{\mathcal F}(Q)
\left(\frac{1}{2}-n_f(p)\right)
\frac{{\mathcal D}_{ik}^q ({\mathbf p}+{\mathbf q})}
{\gamma_{{\mathbf p}+{\mathbf q}}},         
\end{eqnarray}
where 
\begin{eqnarray}
{\mathcal B}(Q)& \equiv & \beta_{\rm L}(Q)+ 
               \left(1- \frac{q_0^2}{{\mathbf q}^2}\right)
               \beta_{\rm T}(Q)  \, ,  \\ 
{\mathcal F}(Q)&\equiv&  \beta_+(Q)\left(1-\frac{q^0}{q}\right) 
                   + \beta_-(Q)\left(1+\frac{q^0}{q}\right)\,. 
\end{eqnarray}
Note that we have approximated the occupation numbers by their 
expansions up to $O(q^0)$ corrections.  The prefactor $2$ 
in front of the piece $g^2 N_f C(r)$ in the last term
of~(\ref{eq:tgluon}) is due to the two possible orientations for the
momentum in the quark loop. The remainder of the previous equations is almost
obvious, after performing the approximations (\ref{eq:fdeltasoft}) and
(\ref{eq:bdeltasoft}).

Finally, as for the case of electrical conductivity, in order to make
more clear the closed resemblance of these equations with Boltzmann-type
equations, we may define the quantities which will turn to  correspond to the
deviations from the equilibrium distribution functions of quarks and gluons, 
\begin{eqnarray}
    \chi_{ik}^q({\mathbf p})&\equiv& \frac{ {\mathcal D}_{ik}^q({\mathbf
    p})}{\gamma_{\mathbf p}} \,, \\
     \chi_{ik}^g({\mathbf p})&\equiv& \frac{ {\mathcal D}_{ik}^g({\mathbf
     p})}{2 |{\mathbf p}|\,\Gamma_{\mathbf p}}\,.
\end{eqnarray}
Then, the substitution of thermal widths by their integral
representations yields the final form of the soft contribution to the
coupled transport equations for
these quantities. % which are completely similar to Boltzmann equations. 
They become
\begin{eqnarray}\label{eq:fvis} 
&&|{\mathbf p}|(\hat{p}^i \hat{p}^k -
\frac{1}{3}\delta^{ik})= \nonumber \\*  
&& \qquad g^2 C_2(r) \int\frac{d^4Q}{(2\pi)^4}\, \pi
\delta(q^0-\hat{{\mathbf p}}\cdot{\mathbf q})\, 
{\mathcal B}(Q) \left(\frac{T}{q^0}-\frac{1}{2}+n_f(p) \right)
  \left[\chi_{ik}^q ({\mathbf p})- 
  \chi_{ik}^q ({\mathbf p}+{\mathbf q})\right] \nonumber \\*
  && \quad + \frac{g^2 C_2(r)}{2 |{\mathbf p}|}
  \left( \frac{1}{2}+n_b(p)\right) \int\frac{d^4Q}{(2\pi)^4}\, \pi
  \delta(q^0-\hat{{\mathbf p}}\cdot{\mathbf q}) {\mathcal F}(Q) 
  \left[\chi_{ik}^q ({\mathbf p})- \chi_{ik}^g
  ({\mathbf p}+{\mathbf q})\right]\,,
\end{eqnarray}
and 
\begin{eqnarray}\label{eq:bvis}
&&|{\mathbf p}|(\hat{p}^i\hat{p}^k - \frac{1}{3}\delta^{ik})= \nonumber \\ *
&& \qquad g^2 N_{c}
\int\frac{d^4Q}{(2\pi)^4}\, \pi \delta(q^0-\hat{{\mathbf p}}
 \cdot{\mathbf q})\, {\mathcal B}(Q) 
 \left(\frac{T}{q^0}-\frac{1}{2}-n_b(p)\right)
 \left[\chi_{ik}^g ({\mathbf p})- 
  \chi_{ik}^g ({\mathbf p}+{\mathbf q})\right] \nonumber \\*
   && \quad+ \frac{g^2 N_{f} C_2(r)}{|{\mathbf p}|} \left( \frac{1}{2}-
  n_f(p)\right) \int\frac{d^4Q}{(2\pi)^4} \pi
  \delta(q^0-\hat{{\mathbf p}}\cdot{\mathbf q})\,{\mathcal F}(Q)
   \left[\chi_{ik}^g ({\mathbf p})- \chi_{ik}^q
  ({\mathbf p}+{\mathbf q})\right]\,.
\end{eqnarray} 

%%%%%%%%%%%%%%%%%%%%%%%%%%%%%%%%%%%%%%%%%%%%%%%%%%%%%%%%%%%%%%%%%%%%%%%%%%%%%%%%%%%%%%%%%

\section{\label{sec:log}Extracting the leading-log transport 
coefficients from the soft contribution}

Rotational invariance fixes the form of the $\chi$-functions.  
In the case of electrical conductivity, they have the form 
\begin{equation}
    \chi_{i}^{s^-}({\mathbf p})=\hat{p}^i \chi^{s^-}(p), 
\end{equation}
and for shear viscosity, 
\begin{equation}
    \chi_{ik}^{q,\, g}({\mathbf p})=
    \left(\hat{p}^i \hat{p}^k -\frac{1}{3}\delta^{ik}\right)
    \chi^{q,\, g}(p). 
\end{equation}

When $|{\mathbf q}| \ll|{\mathbf p}|$,  we may to approximate the
integral collision terms in the transport equations by a derivative expansion
of the $\chi$-functions, in a similar way to the
procedure which leads to the derivation of Fokker-Planck equations in
the context of classical plasmas~\cite{Gasiorowicz,Rosenbluth}.  With the aim 
to extract
the leading-log terms for the transport coefficients, we only 
require the insertion into the transport equations of the second-order
approximations for the quantities
$\hat{p}^i\left[\chi_{i}^{s^-}({\mathbf p})-\chi_{i}^{s^-}({\mathbf
p}+{\mathbf q})\right]$ and $\hat{p}^i
\hat{p}^k\left[\chi_{ik}^{q,\,g}({\mathbf
p})-\chi_{ik}^{q,\,g}({\mathbf p}+{\mathbf q})\right]$.  At
the same accuracy, terms of the type $\left[\chi_{ik}^{q,\,g}({\mathbf
p})-\chi_{ik}^{g,\,q}({\mathbf p}+{\mathbf q})\right]$ are replaced by
$\left[\chi_{ik}^{q,\,g}({\mathbf p})-\chi_{ik}^{g,\,q}({\mathbf
p})\right]$.  These substitutions produce a combination of derivatives 
of the $\chi$-functions, whose coefficients are integral expressions involving  
the Landau damping piece of soft spectral densities.  Here, the
leading-log terms will arise from to the logarithmic dependence on the
upper limit $q_{c}$.  Although the computation of the complete leading
order in $g$ of these integrals can need numerical quadrature, the
leading-log order is easily obtained analytically by means of the
formulas based on sum rules   
given in the appendix~\ref{app:srule}.  

After the angular integration of the Eqs.~(\ref{eq:fvis}) and
(\ref{eq:bvis}), and the subsequent $q^0$ integration with the aid of
Eqs.~(\ref{eq:sumlong}) and~(\ref{eq:sumtrans}), 
the logarithmic terms finally combine to give the
following expressions in the case of shear viscosity
\begin{eqnarray}\label{eq:flog} 
p&=&-\frac{g^2 C_{2}(r) m_{D}^2 T}{16
\pi}\ln g^{-1}
\left\{\chi^{q}(p)''+\left[\frac{2}{p}-\frac{1}{T}(1-2n_{f}(p) )
\right]\chi^{q}(p)' -\frac{6}{p^2}\chi^{q}(p) \right\} \nonumber \\*
&& 
+\frac{g^2 C_{2}(r) \omega_{0}^2}{8 \pi p}
\ln g^{-1}\left(1+2 n_{b}(p)\right) 
\left[\chi^{q}(p)-\chi^{g}(p)\right], \\*
\label{eq:blog}
p&=&-\frac{g^2 N_{c} m_{D}^2 T}{16\pi}
\ln g^{-1}
\left\{\chi^{g}(p)''+\left[\frac{2}{p}-\frac{1}{T}(1+2n_{b}(p) )
\right]\chi^{g}(p)' -\frac{6}{p^2}\chi^{g}(p) \right\} \nonumber \\*
&& 
+\frac{g^2 N_{f }C(r) \omega_{0}^2}{4 \pi p}
\ln g^{-1}\left(1-2 n_{f}(p)\right) 
\left[-\chi^{q}(p)+\chi^{g}(p)\right] ,
\end{eqnarray}
where $\ln g^{-1}$ is the term coming from 
either $\ln\left(\frac{q_{c}}{m_{D}}\right)$ or 
$\ln\left(\frac{q_{c}}{\omega_0}\right)$ with logarithmic accuracy, and 
\begin{eqnarray}
m_D^2 &=& \frac{g^2 T^2}{3} \left[N_c+N_f C(r) \right],\\*
\omega_0^2 &=& \frac{g^2 T^2}{8}\, C_2(r).
\end{eqnarray} 

Quite remarkably, the terms 
$(1\pm 2 n_{b,\,f}(p))/T$ multiplying the first derivatives and generated by the
zero-order in $q^0$ of $n_{b}(q^0) \mp n_{b\,,f}(p+q^0)$, allow us to 
write a single functional of $\chi^{q}(p)$ and
$\chi^{g}(p)$ whose variation leads to above equations for shear viscosity in
the leading-log approximation.
This functional
$\int_{0}^\infty dp\,{\mathcal L}(\chi,\chi')$ 
turns out to be exactly the same as that previously
found by Arnold, Moore and Yaffe~\cite{Arnold1} in their derivation of the
leading-log terms of the linearized collision integral of the 
Boltzmann equation.  After
multiplication of both sides of Eqs.~(\ref{eq:flog}) and
(\ref{eq:blog}) by $2 d(r) N_{f} p^2 n_{f}(p)(1-n_{f}(p))$ and $ d(G) p^2
n_{b}(p)(1+n_{b}(p))$ respectively, one finds
\begin{eqnarray}\label{eq:varshear}
    {\mathcal L}(\chi,\chi')&=& -2 d(r) N_{f} p^3
n_{f}(p)(1-n_{f}(p)) \chi^q(p) - d(G) p^3 n_{b}(p)(1+n_{b}(p))
\chi^g(p) \nonumber \\
      &&+\frac{g^2 N_{f} \omega_{0}^2 d(r) C_{2}(r) }{8 \pi}
      \ln g^{-1}\, p\, n_{f}(p) (1+n_{b}(p)) 
      \left[\chi^q(p)-\chi^g(p) \right]^2 \nonumber \\ 
      &&+\frac{g^2 N_{f} m_{D}^2 T d(r) C_{2}(r)}{16 \pi}
      \ln g^{-1}\,  n_{f}(p) (1-n_{f}(p))
      \left[ p^2 \chi^q(p)'^2 + 6 \chi^q(p)^2 \right] \nonumber \\ 
      &&+\frac{g^2 N_{c} m_{D}^2 T d(G) }{32 \pi}
      \ln g^{-1}\, n_{b}(p) (1+n_{b}(p)) \left[ p^2
      \chi^g(p)'^2 + 6 \chi^g(p)^2 \right] ,
\end{eqnarray}
where we have used $d(r) C_{2}(r) = d(G) C(r)$ with $d(G)=N_c^2-1$. 

Using the expressions~(\ref{eq:slope}) and (\ref{eq:sv}), and noting that 
$({\hat p}^i{\hat p}^k-\delta^{ik}/3)
 ({\hat p}^i{\hat p}^k-\delta^{ik}/3)=2/3$, we see that the contribution to the shear viscosity 
of each Dirac fermion is 
\begin{equation}
 \frac{\beta}{15\pi^2}\int_0^\infty dp \,p^3
 n_f(p)(1-n_f(p)) \chi^q(p), 
\end{equation}
while each gauge boson contributes as 
\begin{equation}
 \frac{\beta}{30\pi^2}\int_0^\infty dp \,p^3
 n_b(p)(1+n_b(p)) \chi^g(p). 
\end{equation}
Hence, 
the shear viscosity is exactly the value of the functional 
$-\beta/(15 \pi^2)\int_0^\infty dp\, {\mathcal L}$ for the values of $\chi$ 
solving the motion equations.

For the electrical conductivity, the above procedure applied to
Eq.~(\ref{eq:chi}) yields the expression
\begin{eqnarray}\label{eq:fcondlog}
1&=&-\frac{m_{D}^2 T}{16\pi}\ln e^{-1}
\left\{\chi^{s^-}(p)''+\left[\frac{2}{p}-\frac{1}{T}(1-2n_{f}(p) )
\right]\chi^{s^-}(p)' -\frac{2}{p^2}\chi^{s^-}(p) \right\} \nonumber \\
&& 
+\frac{\omega_{s}^2}{8 \pi p}
\ln e^{-1} \left(1+2 n_{b}(p)\right) 
\chi^{s^-}(p),
\end{eqnarray}
where the Debye mass for the photon and the plasma 
frequency for the charged fermionic species $s$ are given by 
\begin{eqnarray}
m_D^2 &=& \frac{e^2 T^2}{3}\left(N_{\rm leptons}+ 
d(r)\sum_{\rm quarks} q_s^2\right)	\\ 
\omega_s^2&=&\frac{q_s^2 e^2 T^2}{8}.
\end{eqnarray}
This equation may be retrieved by varying the functional
$\int_{0}^\infty dp\, {\mathcal L}^s(\chi,\chi')$, where 
${\mathcal L}^s$ may be chosen as
\begin{eqnarray}\label{eq:varcond}
    {\mathcal L}^s &=&-p^2 n_{f}(p)(1-n_{f}(p)) \chi^{s^-}(p)
     +\frac{\omega_{s}^2}{16 \pi}
      \ln e^{-1} p\, n_{f}(p) (1+n_{b}(p))
      \chi^{s^-}(p)^2\nonumber \\
      &&+\frac{m_{D}^2 T}{32 \pi} \ln e^{-1}
      n_{f}(p) (1-n_{f}(p)) \left[ p^2 \chi^{s^-}(p)'^2 + 2 \chi^{s^-}(p)^2 \right].
\end{eqnarray} 
Now, the contribution of each charged Dirac fermion  to the
electrical conductivity is
\begin{equation}
 \frac{\beta}{3\pi^2} \int_0^\infty dp \,p^2
 n_f(p)(1-n_f(p))\chi^{s^-}(p), 
\end{equation}
which implies that this transport coefficient is given by 
$-2\beta/(3\pi^2)\sum_s\int_0^\infty dp\, {\mathcal L}^s$ 
at the stationary values of $\chi^{s^-}$.

This completes the derivation of the leading-log terms for the
transport coefficients.  
The agreement with the
results of Ref.~\cite{Arnold1} is complete when the conversions
$\chi^{q,g} \rightarrow -2 \beta \chi_{\rm AMY}^{q,g}$,
$\chi^{s^-} \rightarrow -2 \beta \chi_{\rm AMY}^{e^+}$ are
performed, and the sum over charged species is restricted to leptons.

%%%%%%%%%%%%%%%%%%%%%%%%%%%%%%%%%%%%%%%%%%%%%%%%%%%%%%%%%%%%%%%%%%%%%%%

\section{\label{sec:end}Conclusion and prospects}

In this paper we have shown how to derive transport equations in
some relativistic many body theories from the summation of uncrossed ladder
diagrams within the imaginary time formalism.  The procedure is
similar to the one used in a quite  remote past by Holstein~\cite{Holstein}
and, for the case of the scalar field theory, yields the correct
results previously derived by Jeon~\cite{Jeon}.  In this treatment, one first
identifies the analytic continuation to real frequencies  for the effective
vertex, 
and then writes the integral equation which it satisfies.  
The two relevant quantities in the vertex equation turn 
out to be the imaginary part of the self-energy 
and the
spectral weight of the ladder rung. 

For the case of gauge theories, we
have derived a couple of transport equations~(\ref{eq:chi}),
(\ref{eq:fvis}) and~(\ref{eq:bvis}) by resummation of the ladder
series, whose graphs are made of the rungs associated with resummed
propagators in the HTL approximation.
%These are the main results of this paper.  
The kernels of the
integral equations derived in this way are the same as  the integrands
of thermal widths for hard particles to leading and next to leading order. 
By extracting the logarithmic terms, the
transport equations turn out to be differential equations similar to
Fokker-Planck equations which appear as approximations to the
collision integrals in Coulomb plasmas.  The 
$O(g^4 \ln1/g)$ thermal widths in these equations are seen as damping
terms, similar to those which arise when the Boltzmann equation is
treated in the relaxation time approximation. 
These differential equations are the same as those recently found by
Arnold, Moore and Yaffe~\cite{Arnold1} by analyzing the infrared
divergences of the linearized collision integrals without screened
interactions.
Thus,  %by one hand, 
this fact constitutes a non trivial check of
the formalism we have used and also of the HTL approximation, because 
of the correct matching of the UV-divergences in our approach with 
the infrared divergences in the approach of Ref.~\cite{Arnold1}. 

With respect to the computation of the complete leading order of the
transport coefficients, we believe that, in analogy with the scalar theory, 
the
introduction of all rungs made of the one-loop four-point functions
may be of interest.   Obviously, most of these rungs will lead to
non divergent results, and only a few of them would yield the infrared
divergences of the hard contribution matching with the logarithmic divergences we
have found here.  To carry out these computations within the framework
we have presented, the classification of the four-point functions and
the corresponding spectral densities would have to be studied.

%%%%%%%%%%%%%%%%%%%%%%%%%%%%%Acknowledgements%%%%%%%%%%%%%%%%%%%%%%%%%%%%%%%%%%%%%%%%%%%%

\begin{acknowledgments}

This work is partially supported by grants from CICYT AEN 99-0315 and 
UPV/EHU 063.310-EB187/98. I thank Jos\'e M. Mart\'\i nez Resco for 
helpful discussions. 

\end{acknowledgments}

%%%%%%%%%%%%%%%%%%%%%%%%%%%%Appendix%%%%%%%%%%%%%%%%%%%%%%%%%%%%%%%%%

\appendix 
 
\section{\label{app:bubble}The spectral density of the rung in $\lambda\phi^4$ theory}
 
In this appendix, the spectral density in Eq.~(\ref{eq:bubble}) is calculated. The 
 product of two free Matsubara propagators is conveniently written as the 
 double spectral representation
 \begin{eqnarray}
 G_0(i\nu_q&+&i\nu_k,{\mathbf q}+{\mathbf k})
 G_0(i\nu_k,{\mathbf k})= \int\int\frac{d\omega_1\,d\omega_2}{(2\pi)^2}
 \frac{1}{i\nu_q-\omega_2} \nonumber \\ 
 && \times \left[\frac{\rho_0(\omega_1+\omega_2, {\mathbf q}+{\mathbf k}) 
                 \rho_0(\omega_1, {\mathbf k})}{i\nu_k-\omega_1} 
  - \frac{\rho_0(\omega_1, {\mathbf q}+{\mathbf k}) 
             \rho_0(\omega_1-\omega_2, {\mathbf k})}
             {i\nu_q+i\nu_k-\omega_1}\right],
 \end{eqnarray}
 where $\rho_0(p^0,{\mathbf p})=2\pi\,{\rm sgn}(p^0)\, 
\delta((p^0)^2-p^2)$. By performing the sum over $\nu_k$, and  
taking the imaginary part of the 
analytic continuation $i\nu_q \rightarrow q^0+i 0^+$, one obtains 
\begin{eqnarray}\label{eq:rho1} 
    \rho(Q)&=&\int \frac{d^3{\mathbf
k}\,d\omega_1}{(2\pi)^4} \, n_b(\omega_1)\nonumber \\
 &&\times\left[ \rho_0(\omega_1+q^0, {\mathbf q}+{\mathbf k}) 
                 \rho_0(\omega_1, {\mathbf k})  - 
                \rho_0(\omega_1, {\mathbf q}+{\mathbf k}) 
             \rho_0(\omega_1-q^0, {\mathbf k}) \right]. 
\end{eqnarray}
The second piece of this integral may be rearranged by a shift of the 
$\omega_1$ integration, $\omega_1 \rightarrow \omega_1 +q^0$. Thus, %the 
%previous expression %reduces to 
%\begin{equation}
%\rho_(Q)=\int \frac{d^3{\mathbf k}\,d\omega_1}{(2\pi)^4} 
%\left[ n_b(\omega_1)-n_b(\omega_1+q^0)\right]
% \rho_0(\omega_1+q^0, {\mathbf q}+{\mathbf k}) 
%                 \rho_0(\omega_1, {\mathbf k}), 
%\end{equation}
%which 
after the use of the parity property $\rho_0(\omega_1+q^0, {\mathbf q}+{\mathbf k})=
-\rho_0(-\omega_1-q^0, -{\mathbf q}-{\mathbf k}))$, and the 
substitution 
\begin{equation}
\rho_0(-k^0-q^0, -{\mathbf q}-{\mathbf k}) = 
\frac{1}{1+n_b(-k^0-q^0)}\, G_0^>(-K-Q), 
\end{equation}
Eq.~(\ref{eq:rho1}) yields the desired result
\begin{equation}
\rho(Q)=(e^{\beta q^0} -1) \int \frac{d^4 K}{(2\pi)^4} 
G_0^>(-K-Q) G_0^>(K). 
\end{equation}

\section{\label{app:srule}Sum rules}

 The required integrals over 
 the Landau damping range of the frequency follow from the sum 
 rules~\cite{Lebellac}
  derived from the analytic properties of the effective propagators. 
 With the notation of Ref.~\cite{Blaizot}, these sum rules are
 \begin{eqnarray}
 \int_{-q}^q\frac{dq^0}{2\pi}\frac{1}{q^0}\,
  {}^\ast\beta_{\rm L}(Q)&=&\frac{m_D^2}{q^2\left(q^2+m_D^2 \right)}- 
  \frac{z_{\rm L}(q)}{\omega_{\rm L}(q)^2},   \\ 
  \int_{-q}^q\frac{dq^0}{2\pi}\,q^0\,
  {}^\ast\beta_{\rm L}(Q)&=&\frac{m_D^2}{3 q^2}- z_{\rm L}(q), \\ 
  \int_{-q}^q\frac{dq^0}{2\pi}(q^0)^3\,
  {}^\ast\beta_{\rm L}(Q)&=&\frac{m_D^2}{5}+\frac{m_D^4}{9 q^2}- 
   z_{\rm L}(q) \omega_{\rm L}(q)^2, \\ 
  %%%%%%%%%%%%%%%% 
   \int_{-q}^q\frac{dq^0}{2\pi}\frac{1}{q^0}\,
  {}^\ast\beta_{\rm T}(Q)&=&\frac{1}{q^2}- 
  \frac{z_{\rm T}(q)}{\omega_{\rm T}(q)^2},  \\ 
  \int_{-q}^q\frac{dq^0}{2\pi}\,q^0\,
  {}^\ast\beta_{\rm T}(Q)&=&1- z_{\rm T}(q), \\ 
  \int_{-q}^q\frac{dq^0}{2\pi}(q^0)^3\,
  {}^\ast\beta_{\rm T}(Q)&=&q^2+\frac{m_D^2}{3}- 
   z_{\rm T}(q) \omega_{\rm T}(q)^2,
 \end{eqnarray}
 where $z_{\rm L,T}(q)$ are the residues at the quasi-particle poles.
 With the aim to extract the logarithmic dependence of the transport equations on $q_c$,
 we use the the approximations valid at large $q$, $q\gg m_D$, 
 \begin{eqnarray}
  z_{\rm L}(q)&=&\frac{8 q^2}{m_D^2}\, \exp\left(-2-\frac{2q^2}{m_D^2}\right), \\ 
  z_{\rm T}(q)&=&1+\frac{3 m_D^2}{4 q^2}\left[
    1+\frac{1}{3}\ln\left( \frac{m_D^2}{8 q^2}\right) \right], \\ 
  \omega_{\rm L}(q)^2&=&q^2 \left[1+ 
    4 \exp\left(-2-\frac{2q^2}{m_D^2}\right)\right], \\
  \omega_{\rm T}(q)^2&=&q^2+\frac{m_D^2}{2}+\frac{m_D^4}{4 q^2}\left[
    1+\frac{1}{2}\ln\left( \frac{m_D^2}{8 q^2}\right) \right], 
 \end{eqnarray}
 and we find 
%Then, we write the right side of the sum rules by retaining 
the quadratic terms in $m_D^2$ which give rise to the logarithmic terms 
$\ln(q_c/m_D)$ 
after the $q$ integration, 
\begin{eqnarray}\label{eq:sumlong}
\int_{-q}^q\frac{dq^0}{2\pi}(q^0)^{2 n-1}
  {}^\ast\beta_{\rm L}(Q)&\approx& 
  \frac{m_D^2}{q^{4-2 n}}\times\left\{
  \begin{array}{ll}
  1  & \quad n=0 \\ 
  1/3 & \quad n=1 \\ 
  1/5 & \quad n=2
  \end{array}
  \right. , \\
  \label{eq:sumtrans} 
 \int_{-q}^q\frac{dq^0}{2\pi}(q^0)^{2 n-1}
  {}^\ast\beta_{\rm T}(Q)&\approx& 
  \frac{m_D^2}{q^{4-2 n}}\times\left\{
  \begin{array}{ll}
  -1/4 & \quad n=0 \\ 
  -3/4 & \quad n=1 \\ 
  -11/12 & \quad n=2
  \end{array}
  \right. .
\end{eqnarray}
The explicit form of the functions $\beta_{\rm L}$ and $\beta_{\rm T}$ is 
\begin{eqnarray}
\beta_{\rm L}(q_0,{\mathbf q})&=& 
\frac{4 \pi m_D^2\, q_0\, q\, \theta(q^2-q_0^2)}{
4 q^2\left\{q^2+m_D^2[1-Q(q_0/q)]\right\}^2+\pi^2 m_D^4 q_0^2}, \\
\beta_{\rm T}(q_0,{\mathbf q})&=& 
\frac{8 \pi m_D^2\, q_0\, q^3 (q^2-q_0^2)\,\theta(q^2-q_0^2)}
{4 q^2\left\{ (q^2-q_0^2)\left[2q^2+m_D^2 Q(q_0/q)\right]+
 m_D^2 q_0^2 \right\}^2 + \pi^2 m_D^4 q_0^2 (q^2-q_0^2)^2} , 
\end{eqnarray}
where the function $Q(x)$ is given by 
\begin{equation}
Q(x)= \frac{x}{2} \ln\frac{1+x}{1-x}\,. 
\end{equation} 
 
\section{\label{app:dampings}Thermal widths}

To leading order in $g$ ($e$), the thermal width of a hard particle is
 obtained by the insertion of one HTL gluon (photon)  propagator into the
 skeleton graph for the one-loop self-energy.  For a hard quark
 (or a charged fermionic species $s$), the resulting expression
 is~\cite{Blaizot}
\begin{equation}\label{damping1}
 \gamma_{p}^{(2)} = g^2 C_{2}(r)\,T \int\frac{d^4Q}{(2\pi)^4}\, \pi
 \delta(q^0-\hat{{\mathbf p}}\cdot{\mathbf q})\frac{1}{q^0}
 \left[\beta_{\rm L}(Q)+ \left(1- \frac{q_0^2}{q^2}\right)
 \beta_{\rm T}(Q)\right] ,   
\end{equation}
with $g^2 C_{2}(r)$ replaced by $q_{s}^2 e^2$ in the case of a
charged fermion.  For a hard gluon, the same expression is valid if 
$C_{2}(r) \rightarrow N_{c}$. 
 
Next, we present the expressions for the thermal widths for hard particles
to order $g^4 \ln g^{-1}$.  The fermion self-energy is showed in
Fig.~\ref{fig:selffermi}, and is written as
 \begin{equation}
  \Sigma(i \omega_n,{\mathbf p}) = 
  -g^2 C_2(r)\, T \sum_{\nu_n}\int \frac{d^3{\mathbf q}}{(2\pi)^3}\,
  \gamma^\mu\, {}^\ast S(i\nu_n+i \omega_n,{\mathbf p}+{\mathbf q }) 
  \gamma^\nu G_{\mu \nu}(i\nu_n,{\mathbf q }). 
 \end{equation}
Making use of the spectral representation of the soft fermion in the HTL 
approximation, 
\begin{equation}
 {}^\ast S(i \omega_n,{\mathbf q})=
 \int_{-\infty}^\infty \frac{dq^0}{2\pi} 
 \frac{{}^\ast\Delta_+(Q)h_+(\hat{{\mathbf q}})+
  {}^\ast\Delta_-(Q)h_-(\hat{{\mathbf q}})}{q^0-i \omega_n}, 
\end{equation}
with $^\ast\Delta_{\pm}$ verifying the parity properties
\begin{equation}
   {}^\ast\Delta_{\pm}(q^0,{\mathbf q}) = {}^\ast\Delta_{\mp}(-q^0,{\mathbf q}), 
\end{equation}
we may perform the Matsubara sum. After, one may extract the imaginary 
part of the continuation 
 $i \omega_n \rightarrow p^0+i0^+$. 
 When $p^0=p$, the dominant contribution 
 comes from the piece of the integrand which multiplies to 
 $\delta(p^0-q^0- \left|{\mathbf p}-{\mathbf k}\right|)$. The replacement of this 
 delta by $\delta(q^0-\hat{\mathbf p}\cdot {\mathbf q})$, valid for 
 $q\ll p$, selects the Landau damping piece of ${}^\ast\Delta_{\pm}$, and the 
 expansion of the remaining terms to lowest order in $q/p$ yields
\begin{eqnarray}\label{eq:dampingf}
\gamma_{p}^{(4)} &=& -\frac{1}{4 p}\,{\rm{tr}} 
   \left({\slashchar p}\,{\rm Im}\,
    \Sigma^{\rm ret}(p^0=p,{\mathbf p}) \right) = 
    -\overline{u}(\hat{\mathbf p},\pm)
    \Sigma^{\rm ret}(p^0=p,{\mathbf p}) 
    u(\hat{\mathbf p},\pm) \nonumber \\ 
    &=&\frac{g^2 C_2(r)}{2 p}\left(\frac{1}{2}+n_{b}(p) \right)
    \int\frac{d^4Q}{(2\pi)^4}\, 
    \pi \delta(q^0-\hat{{\mathbf p}}\cdot {\mathbf q})
    {\mathcal F}(Q), 
\end{eqnarray}
where 
\begin{equation}
{\mathcal F}(Q)\equiv  \beta_+(Q)\left(1-\frac{q^0}{q}\right) 
                   + \beta_-(Q)\left(1+\frac{q^0}{q}\right). 
\end{equation}
The spectral functions  $\beta_{\pm}(Q)$ are given by 
\begin{equation}
\beta_{\pm}(q_0,{\mathbf q})=\frac{\pi \omega_0^2 q_0^2(q \mp q_0) 
\theta(q^2-q_0^2)}{
\left[\pm q_0 q (q \mp q_0)+ \omega_0^2\left\{ 
\pm q_0 [1-Q(q_0/q)] + q\, Q(q_0/q) \right\} \right]^2+
\frac{\pi^2 \omega_0^4 q_0^2(q \mp q_0)^2}{4 q^2}},
\end{equation}
where the frequency plasma  for the fermion is $\omega_0^2= g^2 C_2(r) T^2/8$ or 
$q_s^2 e^2 T^2/8$.
% and the function $Q(x)$ is 
%\begin{equation} 
%Q(x)=\frac{x}{2} \ln\frac{1+x}{1-x}\,. 
%\end{equation} 
The integral~(\ref{eq:dampingf}) has been treated in refs.~
\cite{Arnold3,Kapusta,Aurenche} and, for $q_c \gg \omega_s$, gives
\begin{equation}
\int\frac{d^4Q}{(2\pi)^4}\, 
    \pi \delta(q^0-\hat{{\mathbf p}}\cdot {\mathbf q})
    {\mathcal F}(Q) = \frac{\omega_0^2}{2 \pi}\left[
    \ln\left(\frac{q_c}{\sqrt{2}\omega_s}\right)-1+\ln 2\right]. 
\end{equation}

The $O(g^4 \ln g^{-1})$ thermal width for a hard gauge boson is associated with the 
imaginary part of the self-energy of the diagram in Fig.~\ref{fig:selffermi}. 
It reads 
\begin{equation}
  \Pi_{\mu \nu}(i \omega_n,{\mathbf p}) = 
  2 N_f\, g^2 C(r)\,T \sum_{\nu_n}\int \frac{d^3{\mathbf q}}{(2\pi)^3}\,{\rm tr}
  \left\{
  \gamma_\mu\, S(i\nu_n+i \omega_n,{\mathbf p}+{\mathbf q }) \gamma_\nu
  {}^\ast S(i\nu_n,{\mathbf q }) \right\},
\end{equation}
where the prefactor $2$ comes from the two possible ways to arrange a soft 
fermion propagator in the graph. A similar treatment to the fermionic case 
leads to the result  
\begin{eqnarray}\label{eq:dampingb}
\Gamma_{p}^{(4)} &=& -\frac{1}{2 p}\,
   {\rm Im}\,
    \Pi_{\rm T}^{\rm ret}(p^0=p,{\mathbf p})  \nonumber \\  
    &=&\frac{g^2 N_f C(r)}{p}\left(\frac{1}{2}-n_{f}(p) \right)
    \int\frac{d^4Q}{(2\pi)^4}\, 
    \pi \delta(q^0-\hat{{\mathbf p}}\cdot {\mathbf q})
    {\mathcal F}(Q), 
\end{eqnarray}
where $\Pi_{\rm T}(P)= \Pi_{i j} (\delta^{i j}-\hat{p}^i \hat{p}^j)/2$.
For the case of a hard photon, the correct result is obtained with 
the substitutions 
$g^2 N_f C(r) \rightarrow e^2 \sum_{\rm charged\, species} q_s^2$ and 
$\omega_0 \rightarrow \omega_s$.

%%%%%%%%%%%%%%%%%%%%%%%%%%%%%%%%%%%%%%%%%%%%%%%%%%%%%%%%%%%%%%%%%%%%%

\end{document}